\documentclass[graybox]{svmult}
\usepackage{mathptmx}       
\usepackage{helvet}         
\usepackage{courier}        
\usepackage{type1cm}        
\usepackage{makeidx}         
\usepackage{graphicx}        
\usepackage{multicol}        
\usepackage[bottom]{footmisc}
\usepackage{natbib}
\usepackage{color}
\usepackage{rotating}
\begin{document}
\bibliographystyle{spbasic}

\title*{Introduction to Tidal Streams}
\author{Heidi Jo Newberg}
\institute{Department of Physics, Applied Physics, and Astronomy, Rensselaer Polytechnic Institute, 110 8$^{\rm th}$ St., Troy, NY, 12180, USA, \email{heidi@rpi.edu}
}
%
%
\maketitle

\abstract{Dwarf galaxies that come too close to larger galaxies suffer tidal disruption; the differential gravitational force between one side of the galaxy and the other serves to rip the stars from the dwarf galaxy so that they instead orbit the larger galaxy.  This process produces ``tidal streams" of stars, which can be found in the stellar halo of the Milky Way, as well as in halos of other galaxies.  This chapter provides a general introduction to tidal streams, including the mechanism through which the streams are created, the history of how they were discovered, and the observational techniques by which they can be detected.  In addition, their use in unraveling galaxy formation history and the distribution of dark matter in galaxies is discussed, as is the interaction between these dwarf galaxy satellites and the disk of the larger galaxy.}

\section{An Overview Tidal Streams and Halo Substructure}
\label{sec:1}

The Milky Way is a large, spiral galaxy.  It is composed of stars, gas, and dark matter within its ``halo," which is the region within which the matter is graviationally bound.  Like other similar galaxies in the Universe, the Milky Way is also orbited by gravitationally bound satellites that contain their own stars, gas, and/or dark matter.  Globular clusters contain stars, but are generally free of gas and dark matter.  There are 157 known globular clusters in the Milky Way \citep{2010arXiv1012.3224H}.  Dwarf galaxies contain stars, sometimes gas, and are generally thought to contain large amounts of dark matter (though the dark matter content of dwarf galaxies is debated).  There are about 30 known dwarf galaxies in the Milky Way \citep{2012AJ....144....4M}.  We generally think of globular clusters as one generation of star formation, and dwarf galaxies as more massive satellites with ongoing star formation, but the distinction between these two classes of objects has become more blurred as we find fainter dwarf galaxies and more complex globular clusters.  Dwarf galaxies can contain globular clusters or be orbited by smaller dwarf galaxies.

When a globular cluster or a dwarf galaxy comes too close to the Milky Way, our galaxy's gravitational potential pulls harder on the part of the stellar association that is closer to the Galactic center than on the part that is farther away.  This differential force is called a {\it tidal} force, since it is similar to the mechanism that causes the ocean tides on Earth.  On Earth, it is the gravity of the Moon (and also the Sun) that pulls harder on the oceans that face it than on the oceans on the opposite side.  This difference force stretches the oceans and raises the ocean levels on the sides of the Earth that are towards and away from the Moon.

Tidal forces on stars (or dark matter) in dwarf galaxies and globular clusters pull them off off of their satellite of origin, and into tidal streams that can encircle the Galaxy.  Stars that have been tidally stripped from a satellite of the Milky Way, and that are moving more slowly or are closer to the Galactic center than the satellite's center of mass, will be pulled into lower energy orbits, which are slightly closer to the Galactic center and have shorter orbital periods.  Tidally stripped stars that are moving more quickly or are farther from the Galactic center than the center of mass of the dwarf galaxy or globular cluster will be pulled into higher energy orbits and thus orbit more slowly.  The physics is analogous to that of planets orbiting the Sun; inner planets move faster than outer planets.  At the point that the stars are no longer trapped by the gravity of the dwarf galaxy (or globular cluster), they orbit the Milky Way instead of the center of mass of the satellite.  At that point, they are part of the tidal tail.  As the orbital periods of Milky Way halo satellites are of order hundreds of thousands or a billion years, the timescale over which tidal streams evolve is also billions of years.

The content and density of dwarf galaxies is known to vary as a function of radius.  A typical model for a dwarf galaxy includes ten times more dark matter than baryonic matter.  The dark matter has a much larger scale radius than the baryonic matter, which is more centrally located.  There is typically radial structure in the stellar populations as well; younger, more metal-rich stars are more centrally located than the older stellar populations.  These radial gradients in the content of dwarf galaxies are imprinted on the structure of tidal streams.  

Stars and dark matter whose populations are distributed at larger radius from a dwarf galaxy center will be preferentially stripped from the dwarf galaxy at an earlier time (though there is some mixing of material from all radii in the stripping process).  Material in the tidal streams that is stripped first will generally be located farthest from the dwarf galaxy (though it is not a strict time sequence along the stream because the stripped stars have a distribution of energies at the time they became unbound).  Therefore, we expect the dark matter to be stripped first, followed by the older populations, followed by the younger populations.  This can lead to a trend in stellar populations (and chemical abundance) along the stream.

In addition to radial gradients, dwarf galaxies could have their own possibly complex or asymmetric shapes; they could be rotating, and often contain globular clusters within their gravitational potential.  Each of these more complex attributes could contribute to the structure of the stream.  At least some of the globular clusters in the halo of the Milky Way are believed to have been accreted when dwarf galaxies fell in and then were assimilated (along with their globular clusters) into the stellar halo.  Globular clusters themselves, regardless of their origin, may themselves become tidally disrupted, though until they become unbound to the dwarf galaxy they would be shielded from the tidal forces of the Milky Way.

The length of the tidal tails depends on the mass and scale radius of the satellite (which determine how easy it will be to strip the stars), the path along which the satellite orbits the Milky Way, and the length of time the satellite has been orbiting.  While early papers on tidal streams supposed that all tidal stream progenitors had been orbiting the Milky Way for the age of the Universe, more recent papers recognize that the Milky Way galaxy has likely built up over the age of the Universe through merger processes, and the smaller galaxies that are merging now are likely to have been captured at later times.  Since the tidal forces on a satellite are stronger closer to the Galactic center, more material is stripped from the satellite and acquired by the tidal tails when the satellite is at perigalacticon.  The variable rate of mass loss from the progenitor satellite will also lead to structure in the tidal tails.  Models for generating the tidal stream from the Sagittarius dwarf galaxy, for example, often segregate the tidal stream stars by the perigalactic passage during which they were stripped.

Observed tidal streams of stars could have a known progenitor which still survives, despite the fact that some of its stars have been removed.  Streams might also have no known progenitor either because the satellite from which it formed has been completely disrupted into the tidal stream, or because it has yet to be discovered (for example it might be hidden behind the Galactic disk).  Stars that have been stripped by tidal forces are subsequently incorporated into the stellar halo, while still maintaining some kinematical and chemical memory of their origin.

Before the discovery of tidal streams, the stellar halo was modeled as a smooth distribution of stars, fit by a power law profile ($\rho \sim r^{-\alpha}$, where $\rho$ is the density, $r$ is the distance from the Galactic center, and $\alpha$ is a number close to 3.5; the distribution was often additionally ``squashed," in the sense that the density fell of more rapidly in the direction perpendicular to the Galactic plane).  This was sensible when it was expected that halo stars were formed during a rapid collapse of a monolithic gas could from which the Milky Way was formed \citep[ELS][]{1962ApJ...136..748E}.  It even seemed sensible when the subsequent model of \citet[SZ][]{1978ApJ...225..357S}, which described the buildup of stars in the halo through mergers, was proposed; it was assumed that there were so many individual tidal streams that formed the halo that the overall density was relatively smooth.

We know now that tidal streams cause significant density substructure in galactic halos.  Most of the earlier measurements of the parameters in the original power law models depended on which lumps of tidal debris were probed by the observations.  While it is now clear that at least some of the halo built up through recent mergers, a portion of the halo stars could be part of a smooth density distribution.  If there was an early period of rapid merging, the result would still be a smoothly varying spatial distribution of stars that also appears well-mixed in velocity.  It is not currently known whether {\it all} of the stars in the halo were born in other, smaller galaxies that later merged with the Milky Way, or whether some of the stars were formed during an initial, rapid collapse of the Milky Way, as proposed by ELS in 1962.  Power law models are still fit to the shape of the stellar halo, but a conscious choice must be made about whether to fit the density as a smooth component plus significant overdensities in large tidal streams, or whether to average the halo density over large volumes and thus include the large tidal streams in the fit.

In fact, the majority of the tidal debris that has so far been discovered is between 20 and 50 kpc of the Galactic center.  It may be that interior to the Solar Circle the stellar halo consists of so many tidal streams that have been dispersed over so many orbits that the result is indistinguishable from a rapid collapse.  Farther than 50 kpc, the satellites are subject to very weak tidal forces, very long orbital times, and may be very late to fall into the Milky Way's gravitational well.  Therefore, these very distant satellites might not have formed tidal tails.  Studies indicate that the fraction of halo stars in substructures is a few tens of percent within 30 kpc of the Galactic center, and the fraction increases at larger radii \citep[see][and references therein]{2012ARAA..50..251I}

The Milky Way stellar halo may have a smooth, well-mixed component.  But it also contains dwarf galaxies, globular clusters, tidal streams, and ``clouds."  The ``clouds" are density enhancements in the halo that are many kiloparsecs across and could not possibly be gravitationally bound, but are also not drawn out into linear structures like tidal streams.  These structures are thought to be the result of a minor mergers that come in on highly eccentric orbits, which take them near (or through) the center of the Milky Way.  Simulations show that these mergers could result in a pile-up of stars at the apogalacticon of the orbit, which would look like a large cloud when viewed from inside the Milky Way.  

In addition to the Milky Way, tidal streams have also been observed in other galaxies (see, e.g.,  Chapters 8\&9 of this volume).  These streams are most notable in large galaxies close to the Milky Way, such as the Andromeda galaxy (M31; discussed in Chapter~8).  Note that we are using the term ``tidal stream" to mean the stars pulled from a satellite during a minor merger with a much larger galaxy.  This is a distinct phenomenon from the ejection of stars or gas that may occur during a major merger between two galaxies of similar size.  Major mergers can form structures that give a stream-like appearance.  A classic example of this latter type of structure is the the Antennae Galaxies (NGC 4038 and NGC 4039).

\section{The Discovery of Tidal Streams in the Milky Way}
\label{sec:2}

\subsection{Dwarf galaxies and moving groups}

During the last two decades of the 20$^{\rm th}$ century, following the landmark paper of \citet{1978ApJ...225..357S}, the Galactic structure community began to consider the possibility that dwarf galaxy mergers could be shaping the morphology of the Galactic halo.  \citet{1981ApJ...244..912R} observed metal-rich, main sequence A stars near the south Galactic pole.  Because these stars were far from the star-forming regions of the Galactic disk, and main sequence A stars have a limited liftime (and therefore cannot be found far from their place of birth), they concluded that the A stars could have been the result of an encounter between the Milky Way and a dwarf galaxy.  Although twenty years later these stars were identified as more likely blue stragglers in the thick disk \citep{2004AJ....127.3060G}, the paper was important in pushing astronomers to look for halo moving groups, that would result from dwarf galaxy mergers.  At that time, blue straggler stars were thought to exist only in globular clusters, where the high stellar density would result in stellar mergers and/or larger binary fraction.  

It was quite reasonable in the 1980's to assume that all main sequence A stars must be young.  In fact, a decade later, Preston et al. (1994) were still operating under the assumption that blue stragglers were preferentially found in the high stellar density environments of globular clusters; field main sequence stars were still strongly argued to be young.  We now believe that some (or all) blue stragglers are the result of mergers of binary stars.  This process produces a high mass main sequence star long after the initial formation of the stellar population.  This process can occur at any time, in any stellar population, regardless of stellar density.

Though evidence for moving groups of halo stars continued to grow through the late 1980's and early 1990's, the first solid evidence for tidal streams in the Galactic halo was presented by \citet{mmh1996}.  This paper is the third in a series that measures the proper motions and radial velocities of stars near the north Galactic pole, within 8 kpc of the Galactic plane, and finds a retrograde moving group of ~20 stars with a net velocity down towards the Galactic plane.  In addition to this dominant group, they find evidence for additional velocity substructure.  Though this particular moving group has not yet been connected to other tidal substructure, many observations have since shown the same phenomenon; if a small volume of halo stars are studied, significant structure in the velocities of the stars is observed.  At the same time, theoretical studies of the growth of galaxies in cold dark matter (CDM) cosmology concluded that galaxies grew through mergers \citep{1988ApJ...327..507F}.  

The discovery of the Sagittarius dwarf galaxy, which appeared to be in the process of tidal disruption \citep{1994Natur.370..194I}, added to the evidence for hierarchical merging.  Here was a modern-day example of a dwarf galaxy in the process of being assimilated into our galaxy.  Milky Way globular clusters were identified as possibly stripped from the Sagittarius dwarf spheroidal \citep{1995AJ....109.2533D}.  And \citet{1995MNRAS.275..429L} suggested that other halo globular clusters could represent ``ghostly streams" from dwarf galaxy mergers with the Milky Way, though they were not able to to show that any of their suggested streams were ``unequivocally real."

In the last decade of the $20^{\rm th}$ century, people were talking about kinematic discovery of tidal streams in the halo, but there was no expectation that the tidal streams would be detectable in stellar density. There were expected to be a large number of tidal streams, that came in from random directions, and each contributed a very small fraction of the density at any position in the halo.  The tidal debris from dwarf galaxies would be detectable as a moving group in the Galactic halo for more than a gigayear \citep{1995ApJ...451..598J} and would remain on great circles for more than a gigayear \citep{1996ApJ...465..278J}.  Because models of dwarf galaxy tidal disruption showed that the Sagittarius dwarf galaxy could not have survived on its present orbit for the age of the Milky Way, models with mass-to-light ratios of 100 were put forward to explain it's survival until the present day \citep{1997AJ....113..634I}.  Later, it would be suggested that dwarf galaxies could be deflected into their present orbits by other dwarf galaxies or from the large number of dark subhalos expected to be present in the Milky Way's halo, so it was not required for the Sagittarius dwarf in particular to have survived on its current orbit for a substantial fraction of the age of the Universe.

It should be noted here there there is a vast literature on galaxy mergers in the second half of the twentieth century, and it is a bit difficult to define when that community started to recognize halo substructures that are similar to the Sagittarius dwarf tidal stream in our galaxy.  The term ``tidal stream" was used to describe the distortions induced in galaxies during major mergers, rather than the way we are using it in this book: to describe a group of stars that have been torn off of a much smaller satellite during an accretion event.  Conference proceedings showing very faint tidal features in external galaxies in the late 1990's, but we particularly highligh a paper by \citet{1998ApJ...504L..23S}, that was published at about the same time the Sagittarius dwarf and its tidal stream were being discovered in the Milky Way.  In this paper they discovered a dwarf galaxy in NGC 5907 this is of similar size to the Sagittarius dwarf galaxy.  They suggested that this dwarf galaxy could be causing the previously observed warp in H I.  They also identified a faint, elliptical ring that appeared to be the remains of a dwarf galaxy that had been tidally disrupted.

\begin{figure}[!t]
\includegraphics[scale=.5]{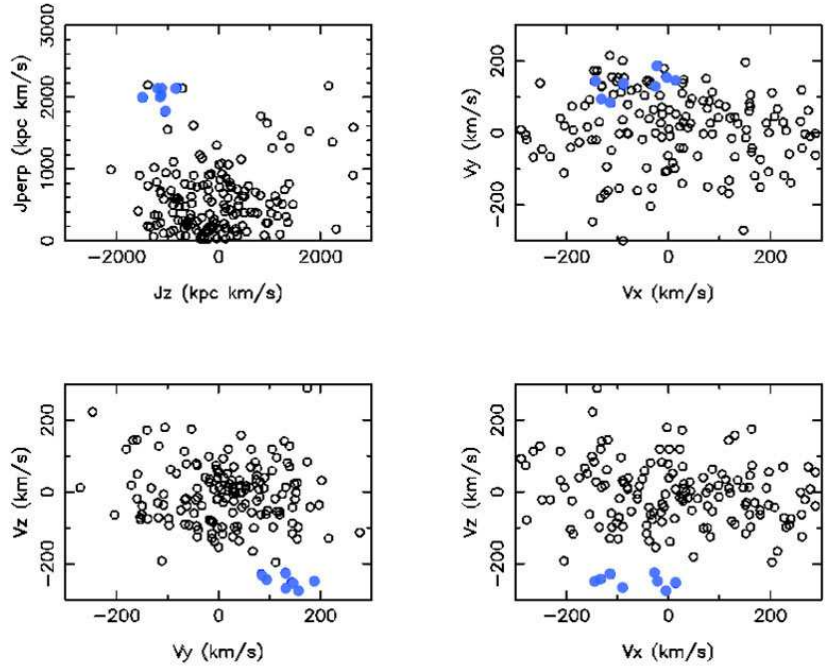}
\caption{These panels show the local velocity distribution and angular momentum ($J_Z$ and $J_{\rm perp}=\sqrt{J_X^2+J_Y^2}$), in Galactic $XYZ$ coordinates, of nearby stars in the Hipparcos survey, following \citet{Helmi1999}.  Here, $X$ points in the direction from the Sun to the Galactic center, $Y$ is in the direction of solar motion, and the coordinate system is right-handed.  The sign of the angular momentum opposite to the original paper, which used a left-handed coordinate system.  The blue filled circles identify stars in a coherent halo moving group.}
\label{hw1999}
\end{figure}

\citet{Helmi1999} successfully identified kinematic substructure in the stellar halo near the Sun, where accurate 3D velocities were available.  They examined the 3D kinematics of 97 low metallicity giant and RR Lyrae stars within 1 kpc of the Sun, and found that about 10\% of them had similar velocities.  We show a re-reduction of Hipparcos catalog data used in that paper in Figure~\ref{hw1999}.  Helmi et al. concluded that 10\% of all metal-poor halo stars outside of the solar circle were accreted from one dwarf galaxy merger with the Milky Way, under the assumption that the local halo represented the halo as a whole.  While the moving group of stars was confirmed by \citet{2000AJ....119.2843C}, the idea that it represented a component of the entire halo was not.  Subsequent papers refer to the moving group as a tidal stream from a satellite that was accreted 6-9 billion years ago, and which comprises about 5\% of the local halo \citep{2007AJ....134.1579K}.

This success in kinematic detection of halo substructure helped to launch the ``Spaghetti Survey" \citep{2000AJ....119.2254M} which aimed to find kinematic substructures through the Milky Way halo using high latitude pencil beam surveys; small fields were imaged in colors that allowed the photometric identification of red giant and blue horizontal branch stars, for which spectra were later obtained.  The final sample \citep{2009ApJ...698..567S} included 101 well-observed halo giant stars, of which 20\% were known to be members of tidal streams.  The survey concluded that between 10\% and 100\% of halo stars are in (presumably accreted) substructure.

\begin{figure}[!t]
\begin{center}
\includegraphics[height=0.9\textwidth, angle=-90]{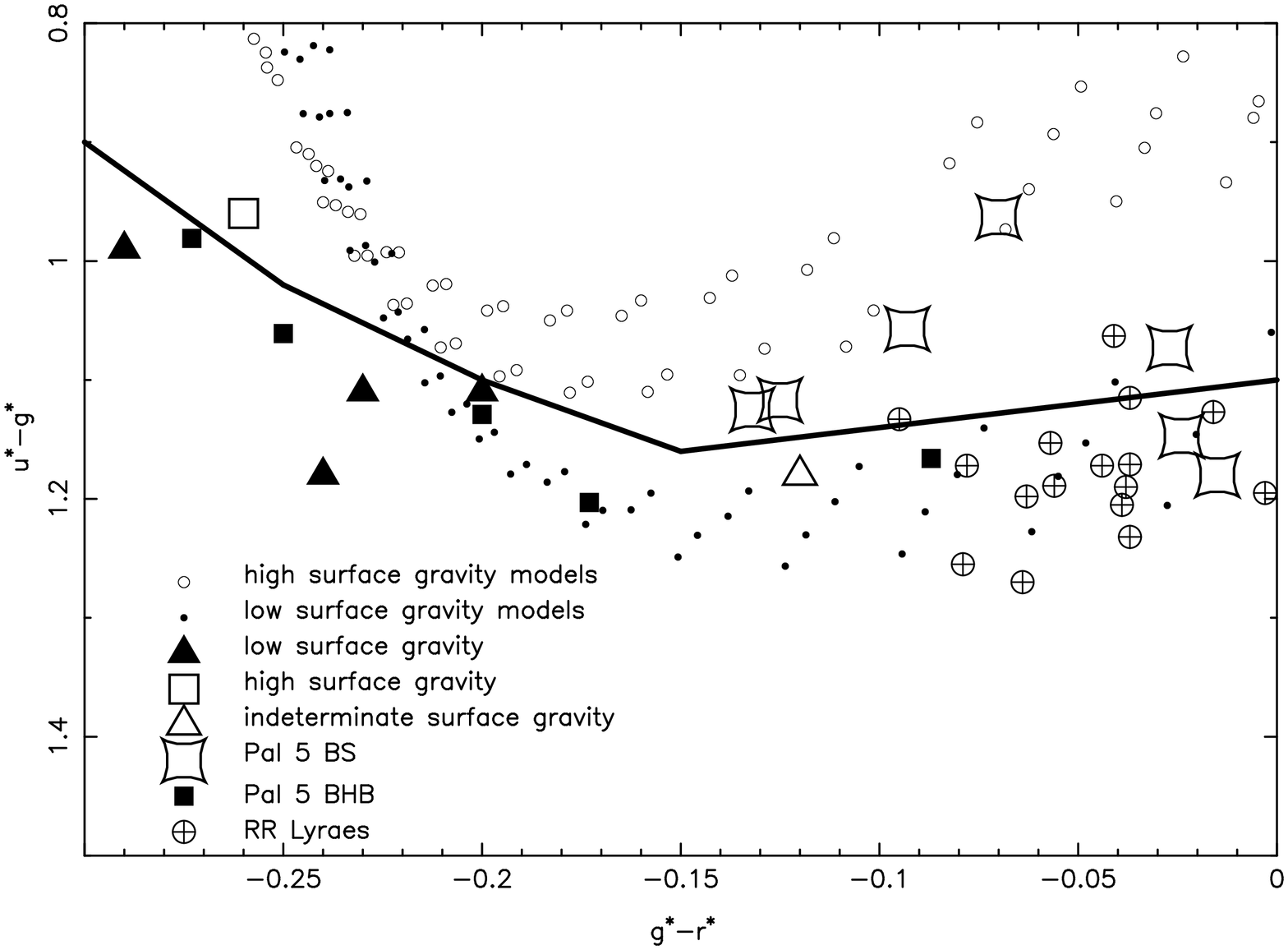}
\end{center}
\caption{This plot shows how low surface gravity BHB stars can be separated from high surface gravity BS stars using $ugr$ colors.  The colors of model stars, with varying metallicity, are from \citet{1998ApJS..119..121L}.  Also shown are colors of BHB and BS stars in Pal 5, and several other stars with spectroscopically determined surface gravities.  The heavy black line is the empirical curve that roughly separates BHBs and BSs by color.  Figure 10 of \citet{2000ApJ...540..825Y}.}
\label{abhbsep}
\end{figure}

\begin{figure}[!t]
\begin{center}
\centering
\includegraphics[height=0.475\textwidth, angle=-90]{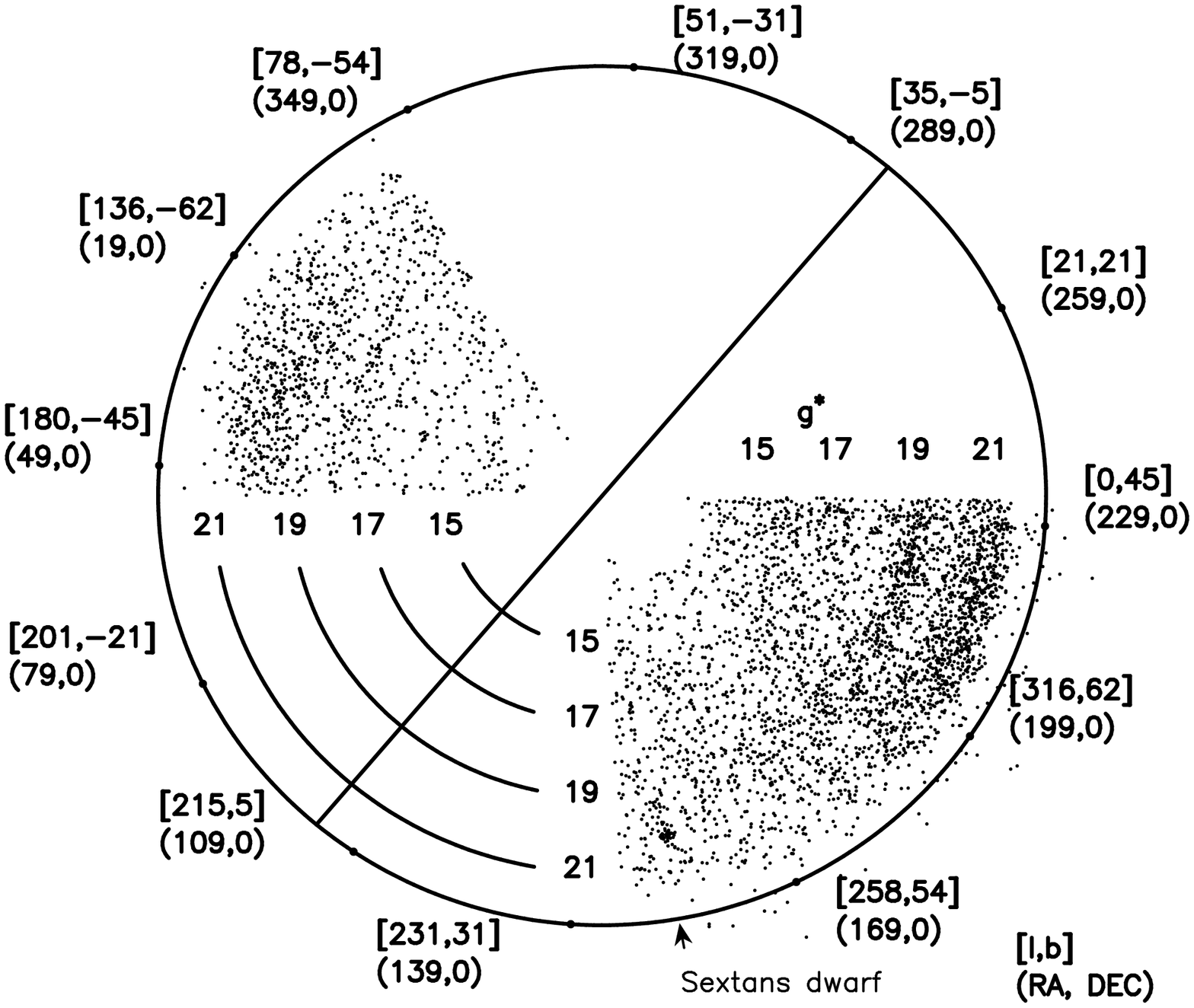}
\par
\includegraphics[height=0.475\textwidth, angle=-90]{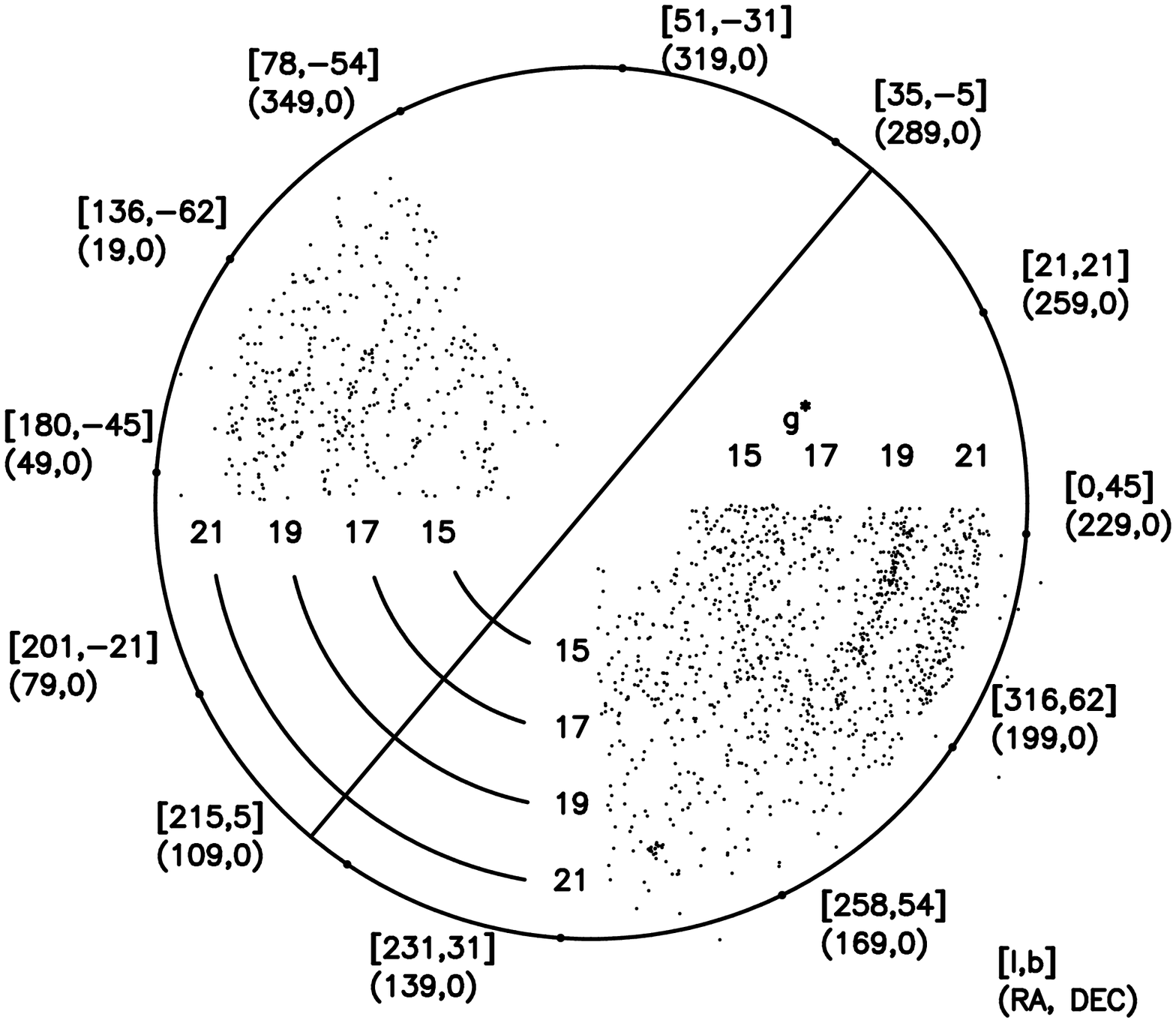}
\includegraphics[height=0.475\textwidth, angle=-90]{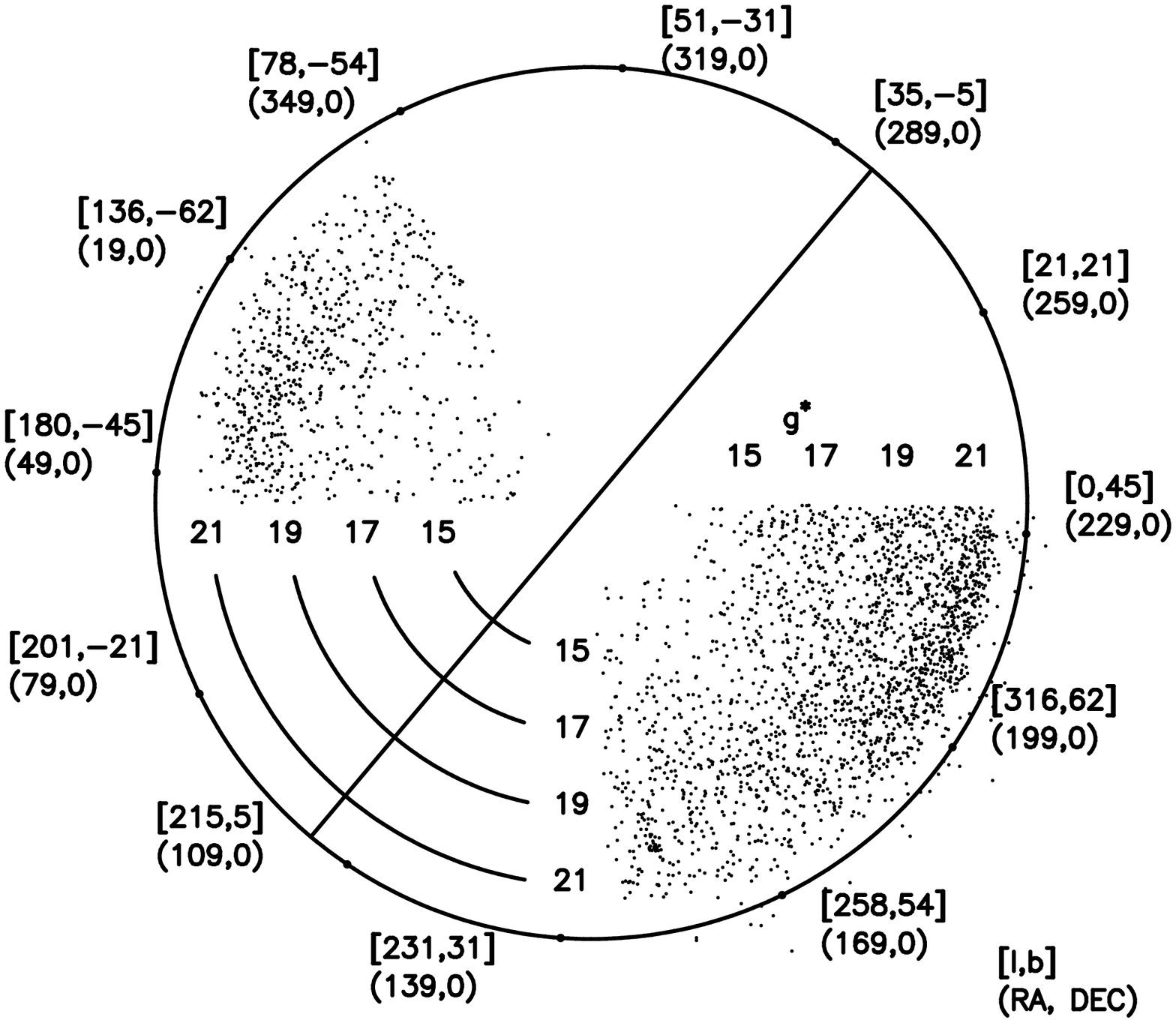}
\end{center}
\caption{These three wedge plots show the blue straggler (BS) and blue horizontal branch (BHB) stars along the Celestial Equator, as selected from early data from the SDSS.  Galactic coordinates are in square brackets and Equatorial coordinates are in parentheses; the solid line shows the intersection of the Celestial Equator and the Galactic plane.  The top plot shows all stars with dereddened colors $0.8 < u^*-g^* < 1.5$ and $-0.3 < g^*-r^* < 0.0$; the lower plots show BHBs (left) and BSs (right), selected by color as shown in Figure~\ref{abhbsep}.  The denser arcs at $195^\circ<\alpha<230^\circ$ and $20^\circ<\alpha<40^\circ$ show the positions where the Sagittarius dwarf tidal stream crosses through the Celestial Equator.  The fainter arcs (larger $g$ magnitude) are the intrisically fainter BS stars, and the same distance as the BHB stars, which are about two magnitudes brighter.  Notice that the BHB stars in the southern (upper left) part of the Sagittarius stream appear to be offset in angle from the BS stars.  It turns out that the Cetus Polar Stream \citep[CPS][]{2009ApJ...700L..61N}, which is very rich in BHB stars, happens to cross through the Celestial equator at almost the same position as the Sgr dwarf tidal stream, and most of the southern BHBs that we originally thought were part of the Sgr stream are actually from the CPS.  Figures 3, 11, and 12 from \citet{2000ApJ...540..825Y}.}
\label{astars}
\end{figure}

\subsection{Halo substructure in the era of large surveys}

The Sloan Digital Sky Survey (SDSS; York et al. 2000) was designed to map the large scale structure of the Universe by measuring the positions of galaxies.  The distances to about one million of the brightest galaxies in 10,000 square degrees would be measured from spectroscopically determined redshifts.  The brightest galaxies would be selected from a high accuracy imaging survey of of the sky in five optical passbands: $u, g, r, i$, and $z$.  Although the extragalactic program provided the primary science driver for the survey, the project provided tremendous opportunities for all other areas of astronomy, and was particularly influential in fostering the field of tidal streams.

First light for the SDSS was obtained in May of 1998, when the imaging survey began with observations of a $2.5^\circ$-wide stripe of data along the Celestial Equator (since observations on the equator did not require the telescope to track).  Unexpected density substructure in the Galactic halo was apparent from the very beginning, though because these were early data reductions it took us a some time to believe that what we were seeing was not an observational artifact.  

\begin{figure}[!t]
\begin{center}
\includegraphics[width=0.8\textwidth, angle=0]{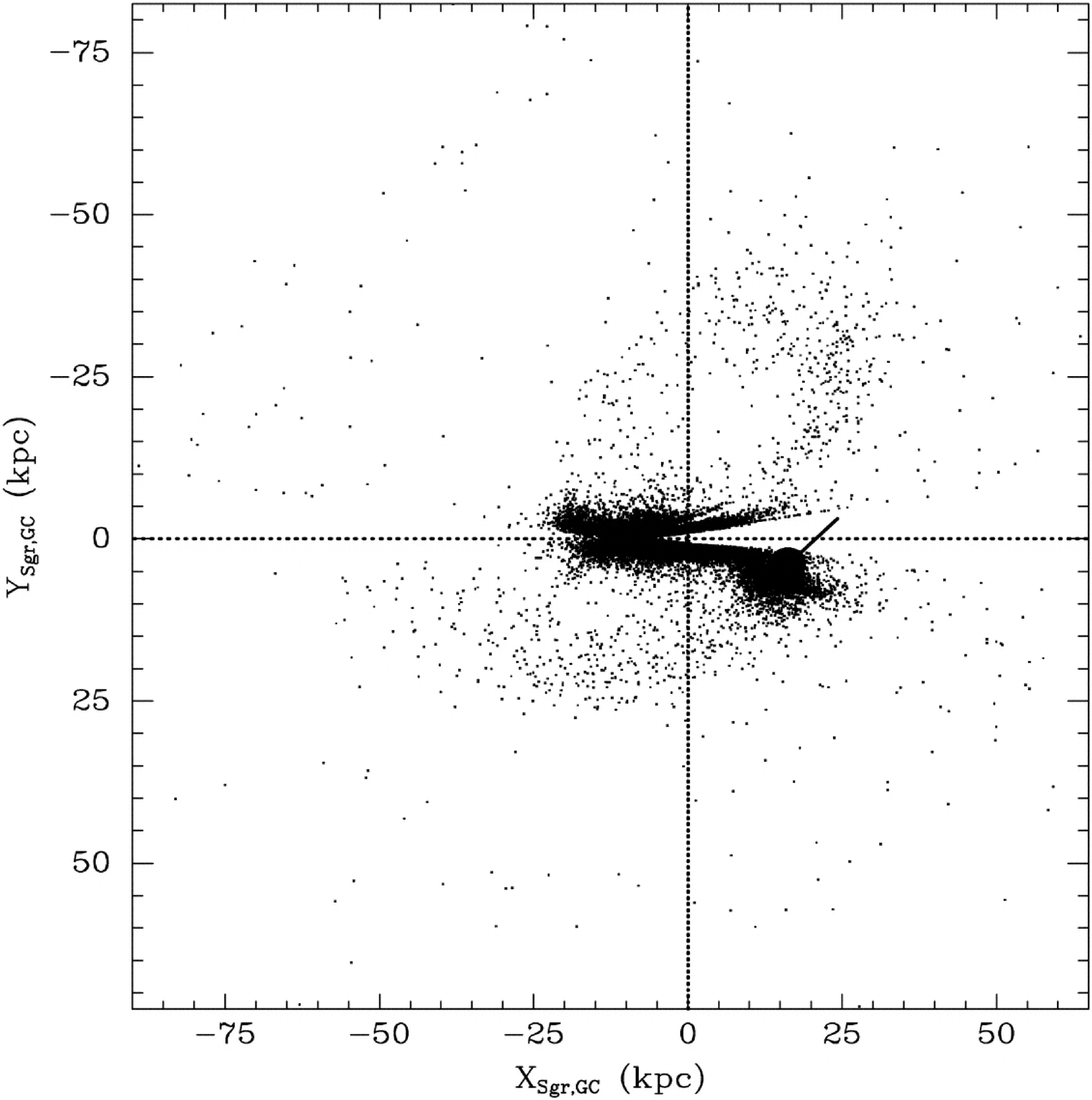}
\end{center}
\caption{The positions of M giant stars within $10^\circ$ of the Sgr dwarf tidal stream, selected from 2MASS, are shown.  The $(X_{Sgr,GC},Y_{Sgr,GC})$ coordinates are the Sagittarius dwarf stream coordinates, where the stream is in the plane $Z_{Sgr,GC}=0$.  Because the Sgr stream is close to polar, this plot is roughly equivalent to Galactic $Z$ vs. $X$, where the center of the Galaxy is at $(0,0)$, the Sun is at $X=-8.5$, $Y$ is the direction of solar motion, and $Z$ points towards the north Galactic pole.  The Sgr dwarf is shown by a large filled circle and a line indicating the direction of its motion.  The M giants trace the path of the Sgr dwarf tidal stream all the way around the Galaxy.  Figure 11 from \citet{2003ApJ...599.1082M}.}
\label{SgrMgiant}
\end{figure}

The first (gravitationally unbound) halo substructure in density was found from blue horizontal branch and blue straggler stars in the first $2.5^\circ$-wide stripe of SDSS imaging data on the Celestial Equator.  \citet{2000ApJ...540..825Y} selected blue stars ($-0.3<g-r<0.0$), that were initially expected to consist entireley of blue horizontal branch stars; main sequence A stars and blue stragglers were generally thought to be absent in field halo populations.  This paper showed that there were two large over-densities of stars - one in the north Galactic cap, and one in the south.  It was also shown that the overdensities included both blue horizontal branch stars and blue stragglers.  The blue stragglers were two magnitudes fainter, and could be roughly separated from the BHB population with $ugr$ colors (Figures~\ref{abhbsep} and \ref{astars}).  The structure in the north Galactic cap was also discovered in SDSS RR Lyrae stars at about the same time \citep{2000AJ....120..963I}.  Because these substructures were diffuse and unlikely to gravitationally bound, we assumed they were two separate, tidally disrupting structures in the Galactic halo.  

It turned out, though, that these two apparently separate structures are two cross sections through the Sagittarius dwarf tidal stream, which is the most prominent tidal stream in the Milky Way.  This was shown by \citet{2001ApJ...547L.133I}, who traced the tidal stream across the sky with carbon stars.  Two years later, this tidal stream was traced across the sky in M giant stars identified photometrically in the Two Micron All-Sky Survey (Figure~\ref{SgrMgiant}, Majewski et al. 2003).  The discovery of a large tidal stream in the Milky Way galaxy was quickly followed up by the discovery of a large tidal stream in the Andromeda galaxy \citet{2001Natur.412...49I}, the Milky Way's nearest neighbor.  

Thus was born the field of tidal streams.  Although the Sagittarius dwarf galaxy tidal stream remains the most prominent and well-studied tidal stream in the Milky Way, even fifteen years later we are still struggling to model the tidal disruption of this dwarf galaxy.  At first, many researchers assumed that the tidal stream from the Sagittarius dwarf galaxy was an anomoly - that this was the only tidal stream that would be identified as a stellar overdensity rather than a co-moving group of stars.  The discovery of density substructure was, after all, unexpected.

It was not long, however, before additional substructure began to be discovered.  \citet{2001ApJ...554L..33V} found an overdensity of RR Lyrae stars in the Virgo constellation.  Tidal tails were found extending tens of degrees from the Pal 5 globular cluster \citep{2001ApJ...548L.165O,2002AJ....124..349R}.  Previous to these papers, globular cluster ``tidal tails" referred to stars outside the tidal radius of the globular cluster, that were identified as ``wings" in the density profiles \citep{1995AJ....109.2553G}.  

\begin{figure}[!t]
\begin{center}
\includegraphics[width=0.8\textwidth]{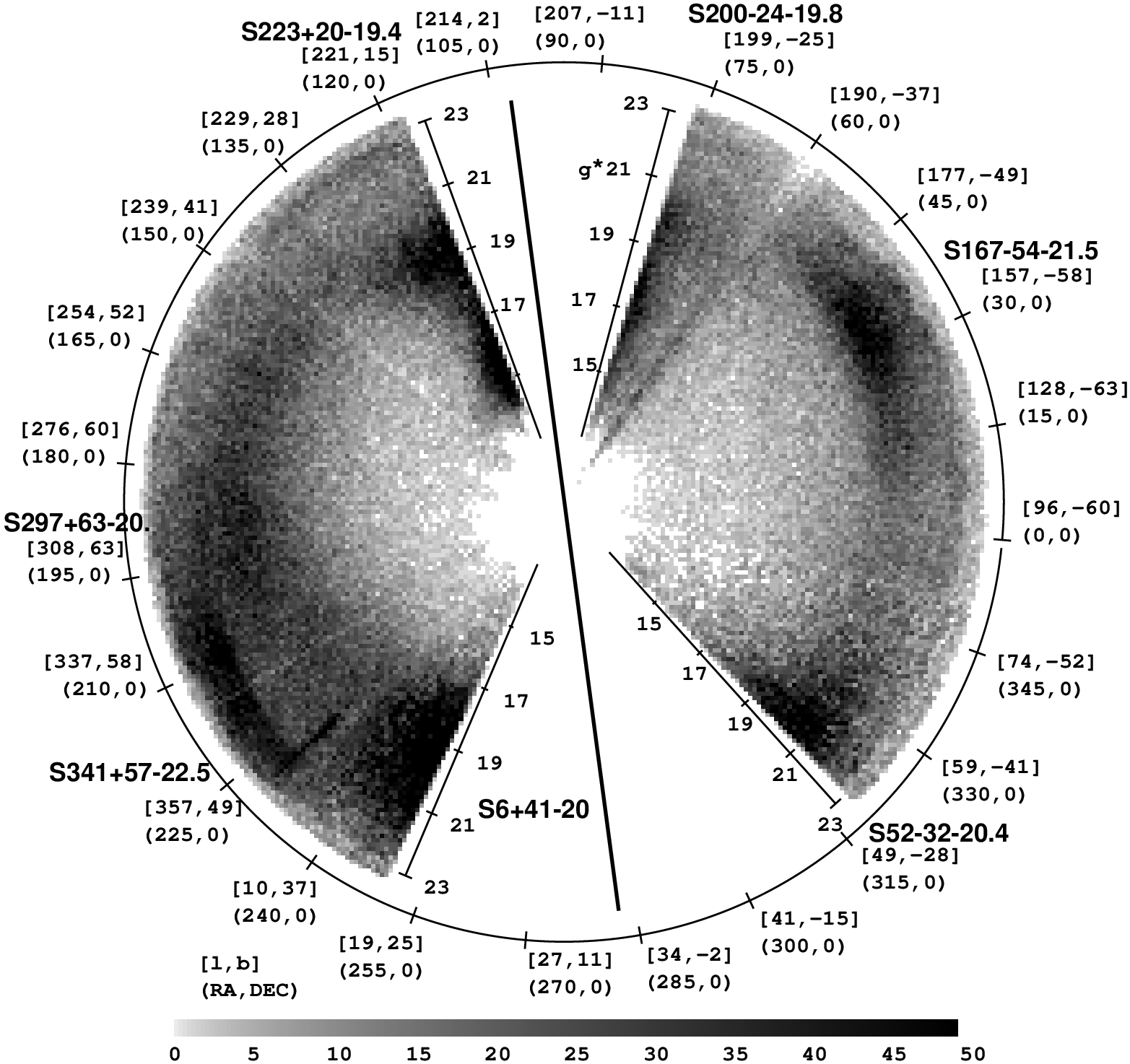}
\vspace{0.25in}
\caption{Wedge plot ($g$ vs. RA) showing the density of 334,066 SDSS turnoff stars $(0.1<(g-r)<0.3)$ on the Celestial equator.  This illustrates the discovery that the Milky Way's stellar halo contains a tremendous amount of density substructure.  The extent and nature of this substructure remains an active area of research, but much of the observed substructure arises from dwarf galaxies that have fallen into the Milky Way and been tidally disrupted.  The angle shows right ascension, and apparent $g^*$ magnitude increases along radius.  The solid line indicates the intersection between the Celestial Equator and the Galactic plane. The Pal 5 globular cluster is evident at $(\alpha,g)=(229^\circ,21)$; the radial extent in the diagram is due to the spread in the absolute magnitudes of turnoff stars.  The Sagittarius stream is at $(l,b,g)=(341^\circ, 57^\circ, 22.5)$ in the north Galactic cap, and at $(l,b,g)=(167^\circ,-54^\circ,21.5)$ in the south Galactic cap.  The Virgo Overdensity is at $(l,b,g)=(297^\circ,63^\circ,20.0)$.  The Hercules-Aquila Cloud juts out from the overdensity at $(l,b,g)=(52^\circ,-32^\circ,20.4)$.  There is a region of high extinction near $\alpha=60^\circ$ that affects the identification of turnoff stars; the radial feature there is not an actual density substructure.  Near the Galactic anticenter, $70^\circ<\alpha<135^\circ$, overdensities can be seen north of the plane at $g=15$ (attributed to the disk in the original paper), south of he plane at $g=17$ \citep[identified as the Monoceros ring in the south by][]{2003MNRAS.340L..21I}, north of the plane at $g=19.4$ (the Monoceros Ring), and south of the plane at $g=20$ (the Triangulum-Andromeda stream, identified as the Monoceros ring in the south in the original paper).  The feature labeled $(l,b,g)=(6^\circ,41^\circ,20)$ is the only overdensity that could plausibly represent the smooth distribution of spheroid stars that had been previously expected.}
\end{center}

\label{wedge}
\end{figure}

In 2002, Newberg et al. made the seminal discovery that the density of main sequence turnoff stars in the Galactic halo is quite lumpy, and is a poor fit to a smooth, power law density profile.  The density of turnoff stars on the Celestial Equator (Figure~\ref{wedge}) includes a wealth of halo substructure.  In this slice through the Celestial Equator, one can see the Sagittarius dwarf tidal stream in both the northern and southern Galactic hemispheres; the Pal 5 globular cluster, which turns out to have long tidal tails; the overdensity in Virgo that was identified by \citet{2001ApJ...554L..33V}; stellar concentrations towards the Galactic center that could be due to a smooth component of the stellar spheroid, but also include stars in what was later named the Hercules-Aquila Cloud \citep{2007ApJ...657L..89B}; a controversial overdensity near the anticenter commonly referred to as the ``Monoceros Ring;" a more distant overdensity south of the Galactic plane called the ``Triangulum-Andromeda Stream" (identified in the original paper as Monoceros in the south); and two concentrations of closer stars near the anticenter (one in the north and one in the south) that appeared be due to the disk but at the time did not fit any disk model.

\begin{figure}[!t]
\begin{center}
\includegraphics[width=0.9\textwidth]{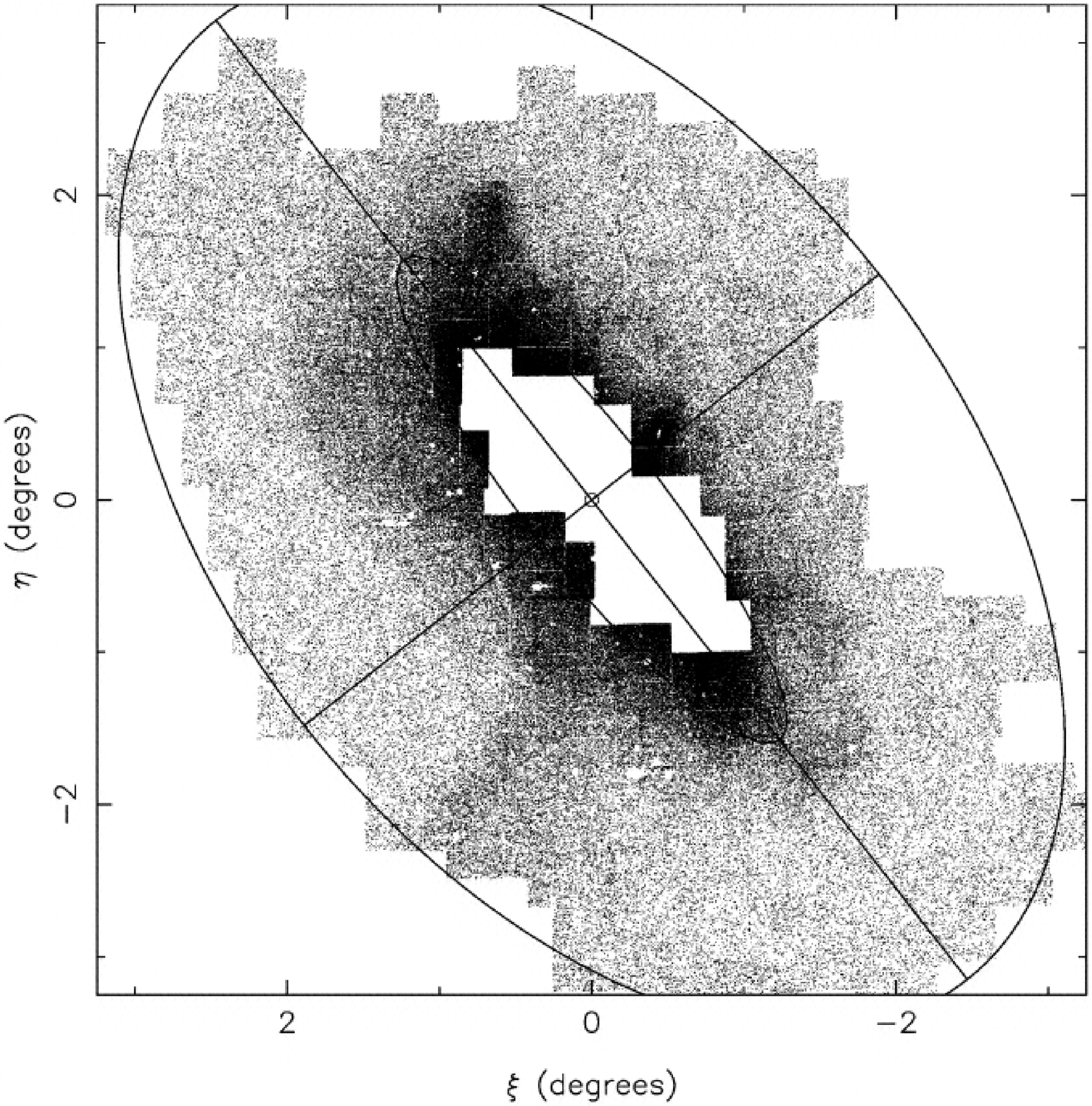}
\caption{We show the surface density of redder red giant branch (RGB) stars in Andromeda (M31).  The outer ellipse has a semimajor axis length of 55 kpc; the inner ellipse has a semimajor axis of 27 kpc and is well outside of the region containing the optical disk of M31.  Halo substructure, including the giant stellar stream at $(\xi,\eta)=(0.5, -1.5)$ and lumpy structure surrounding the disk.  This image showed that the halo of Andromeda contained significant density substructure, just as is seen in the Milky Way.  Figure 2 from \citet{2002AJ....124.1452F}.}
\label{M31}
\end{center}
\end{figure}

Also in 2002, Ferguson et al. showed that there was spatial and metallicity substructure in the halo and outer disk of M31 (Figure~\ref{M31}), using red giant branch stars in a panoramic photometric survey of this galaxy.  M31 is the closest large, spiral galaxy to the Milky Way, and has provided a complementary view of tidal structure in spiral galaxies.  While the Milky Way allows us to see stream structure in three dimensions, it is much easier to see the overall halo structure in external galaxies.

The discovery of lumpy outer halos containing tidally disrtupting dwarf galaxies in both the Milky Way and Andromeda (M31) bolstered the idea that galaxy formation happens through a series of mergers.  The lumps in the halo are evidence that small mergers continue to happen at late times, and the Sagittarius dwarf galaxy is an example of a merger that is still happening today.  Since these discoveries, many more tidal streams have been discovered in both galaxies, and also in galaxies outside of the local group.  

\begin{figure}[!t]
\includegraphics[width=1.0\textwidth,]{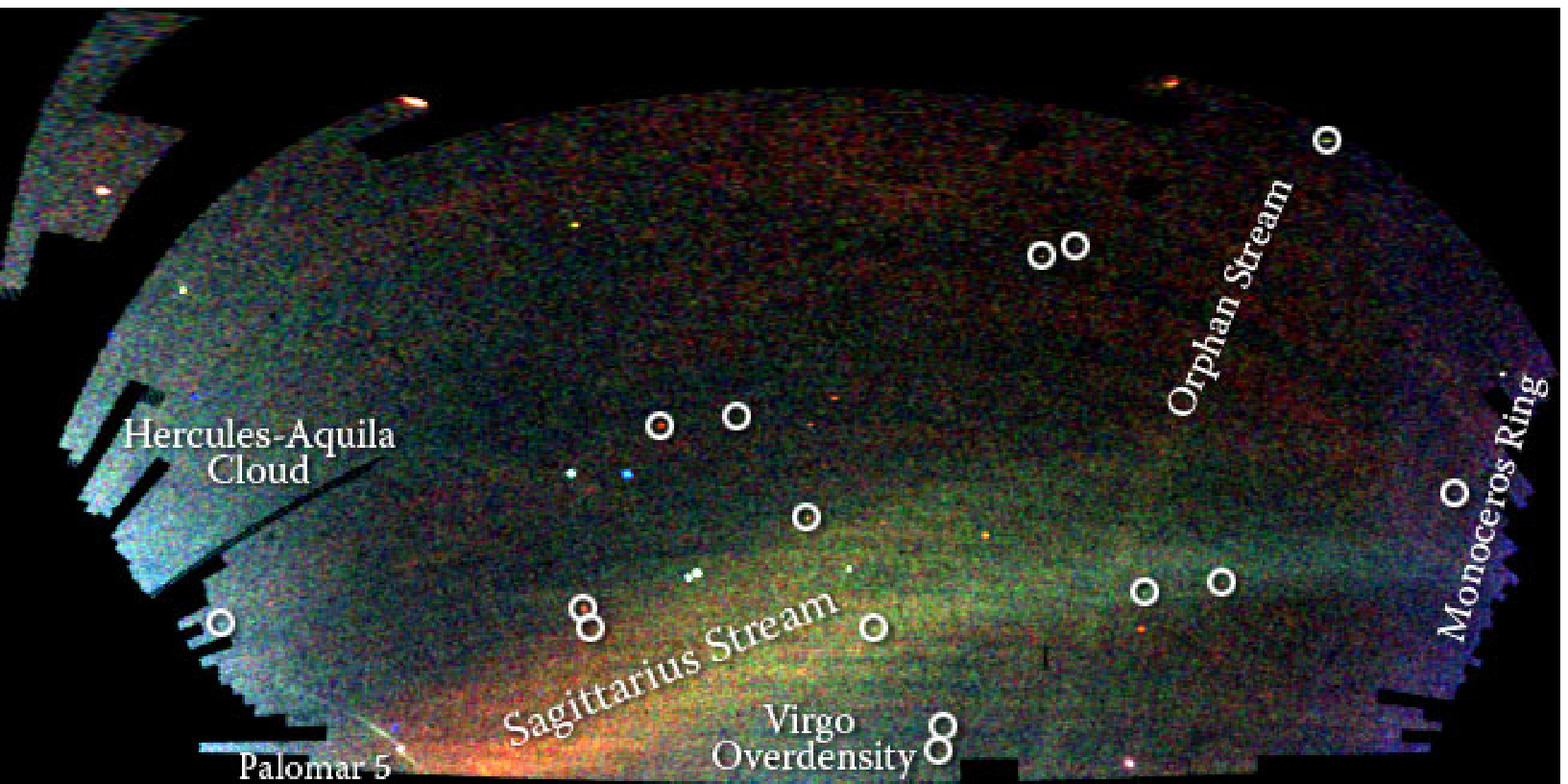}
\caption{This spectacular image of density substructure in the Milky Way's stellar halo is known as the ``Field of Streams."  Declination vs. right ascension (increasing towards the left) is shown for SDSS turnoff stars in the north Galactic cap.  This is an RGB composite where blue represents the brighter (closer than about 15 kpc) stars, and red represents the faintest (more distant than about 25 kpc) stars.  Note that there appear to be two branches to the Sagittarius Stream, and a multitude of halo substructures.  The circles indicate the positions of new Milky Way satellites discovered in SDSS data; two are faint globular clusters and the others are faint dwarf galaxies.  Image credit: V. Belokurov and the Sloan Digital Sky Survey.}
\label{fieldofstreams}
\end{figure}

An amazing map of the density of turnoff stars, color-coded by distance (see Figure~\ref{fieldofstreams}), was published by \citet{fieldofstreams}; it shows the most prominent tidal streams in the north Galactic cap, spread out across the sky.  In this image one can see not only the Sagittarius dwarf tidal stream, but also a fainter tidal stream next to it that may also be associated with Sagittarius, but so far no model has been able to simultaneously fit both of the ``bifurcated" streams.  The Virgo Overdensity is also conspicuous, as is the ``Monoceros Ring," and another smaller tidal stream called the ``Orphan Stream," because it has no known  dwarf galaxy progenitor.  The Orphan Stream progenitor either has yet to be discovered or has been completely disrupted.  

Also apparent, nearer to the Galactic center, is the Hercules-Aquila ``cloud."  Overdensities that are not long and thin like tidal streams, but instead look like large, amorphous clumps are referred to as ``clouds."  These could be produced by tidal debris from a dwarf galaxy on an orbit that takes it near the center of the Milky Way galaxy \citep{1988ApJ...331..682H}.  The tidal disruption is severe as the dwarf galaxy goes through the center of the Milky Way, where the tidal forces are strong.  The stars that have been tidally stripped then orbit back and forth through the Galactic center, spending a larger fraction of their time at apogalacticon, where they are observed as amorphous ``clouds" of stars.  Note, however, that it has been suggested that at least one of these clouds (the Virgo Overdensity) could be the disrupted remains of a dwarf galaxy that is on a highly eccentric orbit, but just recently passed perigalacticon and thus does not consist of debris piled up at the apocenter of its orbit \citep{2012ApJ...753..145C}.

While this section gives an historical account of the major observational milestones in the beginning of the field of tidal streams, the succeeding chapters of this book describe in detail the discoveries that have since been made.  When spatial substructure in the halo was first discovered, Don York, from the University of Chicago, said we had discovered ``Eggen's spaghetti sky."  The same Olin Eggen that is famous for a paper suggesting the Milky Way formed from the rapid gravitational collapse of a cloud of gas \citep{1962ApJ...136..748E} is also famous for seeing clumps of stars in imaging data, and in particular introducing the concept of ``moving groups" of stars \citep{1958MNRAS.118...65E,1958MNRAS.118..154E,1958MNRAS.118..560E, 1959MNRAS.119..255E} - stars that are in a similar volume of space and moving in the same direction.  Although his moving groups are likely from tidal disruption of open clusters in the disk rather than dwarf galaxies and globular clusters the halo, the image of a sky filled with stars from tidally disrupted clusters mixed together like a bowl of spaghetti is a common analogy.  \citet{mmh1996} make a similar analogy to a ``can of worms."  However, the width of these stellar streams is astonishing.  The tidal stream from the Sagittarius dwarf galaxy, for example, has a width of about 6 kpc.  For reference, the distance from the Sun to the Galactic center is about 8 kpc, and the radius of the disk is about 15 kpc.  The width of 6 kpc is a substantial fraction of the width of the Milky Way.  Although tidal streams from globular clusters might create ``spaghetti" in the sky, dwarf galaxies like Sagittarius are made of much thicker pasta!  Perhaps we should call it the ``rigatoni sky!"  This is of course written in jest, but one should not forget the enormous spatial extent of some of these streams and clouds.

\section{Observational Techniques for Finding Substructure}

Tidal streams can either be identified as an overdensity of stars in the halo, or by finding stars with similar locations and velocities (using angular momentum alleviates the need for co-location).  The latter technique will allow the identification of tidal streams with lower surface density, a longer time since disruption of the satellite, or which are closer to the Galactic center where the stellar density is higher.

\subsection{Standard Candles and Photometric Parallax}

The most straightforward way to identify tidal streams by stellar density is to determine the three dimensional positions of halo stars and then look for localized regions that have densities higher than their surroundings.  Measuring positions in 3D requires that one know the distance to each star, so particular stars with known absolute magnitude (standard candles) must be selected.  Often, these standard candles are selected photometrically based on their colors, but they might also be selected from spectroscopy or as variable stars.

\begin{figure}[!th]
\sidecaption
\includegraphics[width=0.6\textwidth]{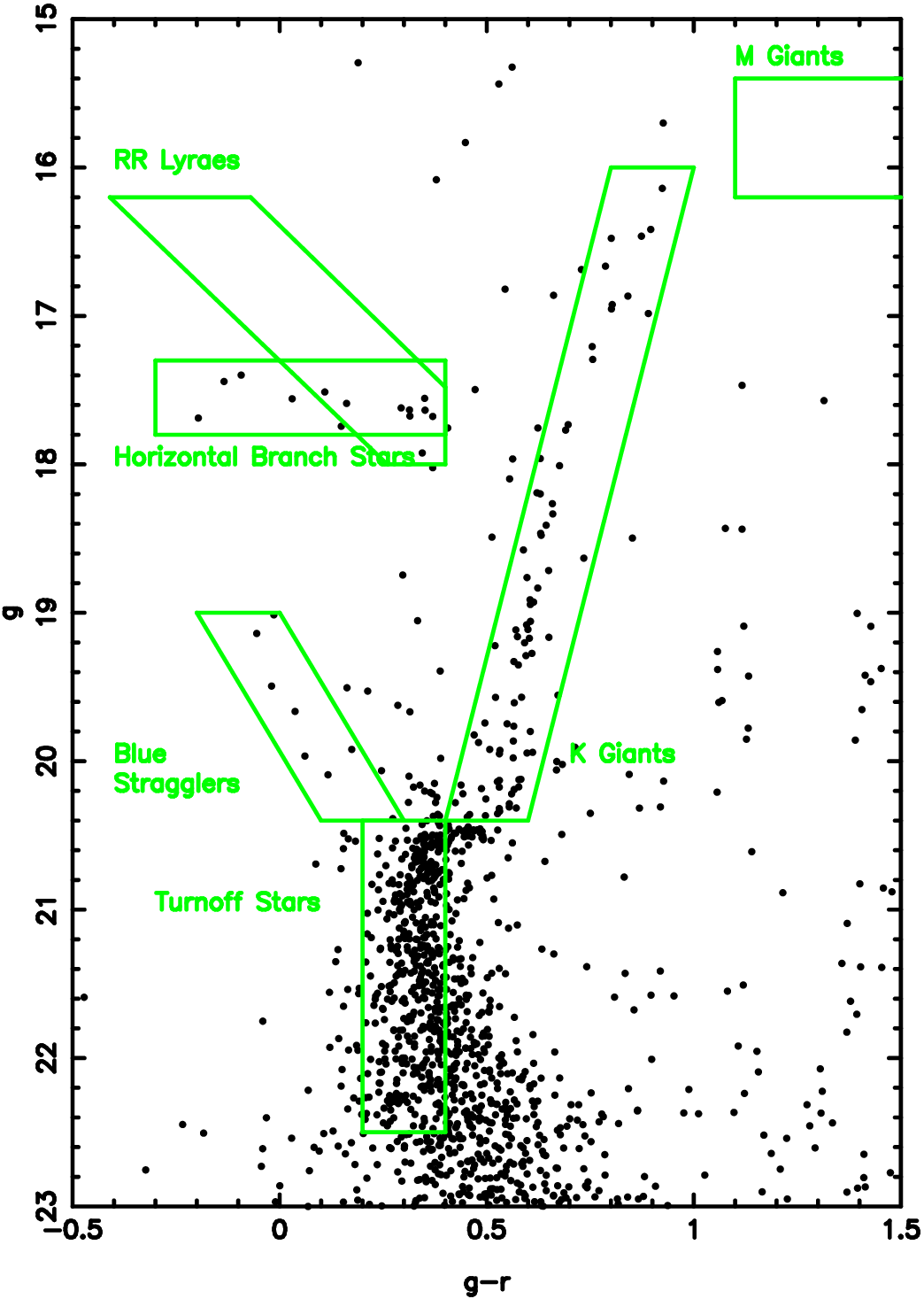}
\caption{The color-magnitude diagram shows stars that can be used to trace Galactic structure.  The black points are SDSS stars within four arcminutes of the Pal 5 globular cluster, and represent the distribution of stars in the halo population.  In a magnitude-limited survey, brighter stars sample the galaxy to larger distances, but they are not very numerous.  Turnoff stars have been popular for tracing halo structure because they are easily identified by color, and are the brightest of the more numerous main sequence stars.  The drawback to turnoff stars is that they have a large spread in absolute magnitude, so individually they are not good standard candles.  This difficulty is somewhat mitigated by the fact that, for halo stars in the Milky Way, the absolute magnitude distribution is not a strong function of population.  Therefore, statistical photometric parallax can be used to recover the underlying density distribution from the positions and apparent magnitudes of turnoff stars.  Figure 1 of \citet{2013IAUS..289...74N}.}
\label{IAUCMD}
\end{figure}

Figure \ref{IAUCMD} shows a color-magnitude diagram (CMD) of stars from the Palomar 5 globular cluster, showing typical features for an old, metal-poor stellar population.  The brighter stars (which allow us to probe the distant halo) include blue horizontal branch stars (BHBs), RR Lyrae variables, red giant branch stars, M giants, blue stragglers, K giant stars, and turnoff stars (note that not all of these types are present in Pal 5!). There are advantages and disadvantages to using each type of star.  

Horizontal branch stars are good standard candles because they occupy a very narrow range of absolute magnitude.   Blue horizontal branch stars (BHBs) can be preferentially selected by blue color, though with some contamination from blue straggler stars which have the surface gravities of main sequence stars.  However, BHBs are only present in very old, metal-poor stellar populations, and there are relatively few stars of this type.  Red horizontal branch stars are more difficult to identify due to confusion with giant branch stars and much fainter main sequence stars.

RR Lyrae stars are even better standard candles, but are harder to identify (usually through multi-epoch variability studies), and also appear only in intermediate-age, fairly metal-poor populations.  RR Lyrae stars are even sparser tracers of the stellar density than horizontal branch stars.  Additional observations (spectra and/or time-series photometry) is required to determine the distance to each RR Lyrae star.  

M giant stars are extremely bright, but only found in metal-rich stellar populations, which are rare in the halo of the Milky Way.  In fact the only verified tidal stream that has been traced with M giant stars is the largest and most apparent overdensity: the Sagittarius dwarf tidal stream (see Chapter 2).  The Monoceros Ring, Triangulum-Andromeda cloud (see Chapter 3), and Pisces overdensity \citep{2010ApJ...722..750S} have also been studied in M giants. 

K giant stars are a natural choice for tracing stellar densities because they are more numerous than other stars of similar intrinsic brightness, and present in all old stellar populations.  The main drawback to using K giant stars is that spectra \citep[or specialized filters that are particularly sensitive to the gravity of K stars; see, e.g.,][]{2000AJ....120.2550M} are required to separate K giants from the much more numerous K dwarf stars in the disk, which have similar apparent magnitudes as K giant stars in the halo. Additionally, the absolute magnitude depends on the age and metallicity, and is a strong function of color (red giant branches are nearly vertical in the CMD); the measured distances will depend heavily on the model used to calibrate them.  

Blue stragglers are best separated from blue horizontal branch stars using spectra, but they can be roughly separated if there is data in the $u$, $g$, and $r$ filters.  These stars are not usually used to trace stellar populations because they are not good standard candles (absolute magnitude is a strong function of color), they are fairly rare, and they are not as bright as giant stars so they do not probe as far in distance.  However, they are often observed in tidal streams, two magnitudes fainter than the blue horizontal branch stars, and must be considered particularly when using BHB stars to trace stellar density.

Turnoff stars are much more numerous than any of the other types of stars considered above, but they are by far the worst standard candles - spanning at least two magnitudes at about constant color.  When using this type of star to measure density along the line-of-sight, it is advantageous to use {\it statistical photometric parallax}, which is described in Section 3.3.

\subsection{Matched Filter Techniques}

Matched filter techniques are used in many applications to optimally extract signal from noise.  In general, one convolves the observed, noisy signal (as a function of either space or time) with a template (filter) that looks like the expected signal.  This will enhance signal and surpress noise.  

In this case, we would like to find the angular density of stars in a tidal stream population (that have a known distribution in color and apparent magnitude), while filtering out stars that are from other disk or halo populations (for which it is also possible to construct an expected distribution in color and apparent magnitude).  

Matched filtering was first used to detect tidal streams by \citet{2002AJ....124..349R}, who discovered the tidal tails of the Pal 5 globular cluster in SDSS data.  The matched filter technique described here follows the description in that paper.  In this version of the matched filter, the cluster stars that are in regions of the color-magnitude diagram where few background stars are expected are weighted more heavily; a {\it minimum variance estimate} (defined below) for the number of cluster stars is derived, rather than a maximum signal-to-noise estimate.  By more heavily weighting the stream detections in regions of low background, systematics from poor knowledge of the background is reduced.

One defines a normalized probability in color-magnitude space that describes where one expects to find stars in the stellar population of the stream (including primarily turnoff, giant, and blue horizontal branch) as $N_{str}$.  In the case of Pal 5, the CMD of the GC itself, normalized so the total probability integrated over the whole color-magnitude diagram is one, is used for $N_{str}$.  In addition, one defines a normalized probability in color-magnitude space for the background, $N_{bg}$.  In general, this background depends on Galactic latitude and longitude.  Then, an estimate for the number of stream stars, $\alpha$, in a given solid angle can be found by minimizing the following expression for $\chi^2$:
\[\chi^2 = \sum\limits_{color, mag}^{}\frac{\{N(color, mag)-[\alpha N_{str}(color, mag) + N_{bg}(color, mag)]\}^2}{N_{bg}(color, mag)},\]
where $N(color, mag)$ is the number of stars in a particular color, magnitude bin.  The sums are over all of the bins that have a non-zero number of stars expected in the stream.  The background must be constructed so that the expected number of stars is also non-zero in all of these bins.  This equation can be solved for the $\alpha$ that corresponds to the minimum $\chi^2$ by setting the first derivative with respect to $\alpha$ equal to zero:
\[\alpha = \frac{ \sum\limits_{color, mag}^{}(\frac{N(color, mag) N_{str}(color, mag)}{N_{bg}(color, mag)} - N_{str}(color, mag)) }{\sum\limits_{color, mag}^{} \frac{N^2_{str}(color, mag)}{N_{bg}(color, mag)}}.\]
In the limit of infinitely small bins, the first term in the numerator can be replaced by a sum of $N_{str}(color, mag)/N_{bg}(color, mag)$ for each of the observed stars in a particular solid angle.  The second term in the numerator and the denominator can  be replaced by an integral over color-magnitude space rather than a sum over the individual bins.

In the case of Pal 5 tidal tails, $N_{str}$ was derived from stars with high membership probability in the globular cluster.  The background distribution, $N_{bg}$, was derived from stars in several square degree patches of actual SDSS data near the globular cluster.  Because the background distribution was derived from the actual survey data, it includes features due to both the survey itself (including increased color errors and decreased completeness near the limiting magnitude of the survey), and the density distribution of stellar populations in the Milky Way.  The tidal tails of Pal~5 were expected to be at about the same distance from the Sun as the cluster itself across the region surveyed, so no shift in apparent magnitudes of stream stars was considered.  Because the parts of the background color-magnitude diagram that varied across the part of the sky surveyed were at colors and magnitudes where there were few stream stars, a varying background distribution was not considered in this paper.  Rockosi et al. constructed a smoothed density plot of the tidal tails of Pal 5 with a resolution of $0.09^\circ$ by finding the optimal value of $\alpha$ as a function of sky position.

Grillmair and Johnson (2006) modified this matched filter technique so that it could be applied to the discovery of tidal streams when the distance to the streams and the stellar population is not known a priori.  The description here follows Grillmair (2009).

The probability distribution for the stream stars is generated from globular cluster stars observed in SDSS data.  The mean $g-r$ and $r-i$ colors are determined for a fixed set of $g$ magnitudes in each of the clusters; colors at other magnitudes are interpolated.  The distribution can be shifted brighter or fainter to account for different distances to the tidal debris.  The relative number of stars expected in the stream as a function of apparent magnitude (the luminosity function) is determined from example globular clusters with very deep photometry available, and modified by the expected completeness as a function of apparent magnitude for the SDSS survey.  It is broadened in color by the amount of the expected photometric error as a function of apparent magnitude.  An additional broadening with $\sigma=0.02$ magnitudes is also applied to account for the intrinsic dispersion in the colors of giant branch stars.

This new technique makes it possible to create a set of 2D pictures of the sky, that show the matched-filter density maps of different stellar populations, at different distances.  These can then be searched for linear structures that indicate the presence of a tidal stream.  One creates ``filters" that consist of the expected color-magnitude distribution of stream stars in the SDSS, at different distances and using different globular cluster distributions as the template.  A variety of background distributions, distances, and stellar populations, are used to create the 2D pictures.  One can scan through a distance series of these pictures, for a fixed stellar population, to pick out linear structures that might vary in distance as a function of position in the sky.

\subsection{Statistical Photometric Parallax}

If one observes objects whose stellar type can be identified from photometry, and which therefore have a known absolute magnitude, then distances to these objects can be determined by {\it photometric parallax} (see section 3.1). An example where photometric parallax for a large number of stars was used to measure the density of stars in the Milky Way is \citet{2008ApJ...673..864J}.  They estimated the distances to 48 million SDSS stars of type G and later, under the assumption that they were main sequence stars (which most of them undoubtedly are), and using a photometric calibration from \citet{2008ApJ...684..287I}.  From the estimates of the individual distances, the positions of each star are derived, and from the resulting stellar densities the structure of the Milky Way (including tidal substructures) was derived.

In contrast, {\it statistical photometric parallax} \citep{2013IAUS..289...74N} uses a statistical approach to determine the underlying spatial density of stars from the distribution of observed apparent magnitudes.  It differs from {\it photometric parallax} because distances are not determined individually for each star.  This technique has been used to determine the spatial distribution of stars in the Sagittarius dwarf tidal stream, using turnoff stars from the SDSS \citep{2013AJ....145..163N}.  Although turnoff stars have a wide variety of absolute magnitudes, and are thus not very good standard candles, they are quite useful for density mapping because they are much more numerous than giant stars, and can be seen at larger distances than fainter, lower mass main sequence stars.  Statistical photometric parallax is most successful for measuring structures with comparable (or higher) stellar density than the background at their location.

The basic concept of statistical photometric parallax is demonstrated in Figure~\ref{wedge}.  There is a narrow, radial feature at RA=$229^\circ$ at an apparent magnitude of $g\sim21$.  This feature is the globular cluster Pal 5, whose stars are all essentially the same distance from the Sun, but the apparent magnitudes of the turnoff stars in the figure span an apparent magnitude range from $g\sim20$ to $g\sim22.5$.  The large range of apparent magnitude is a result of the large range of absolute magnitudes of turnoff stars.  In fact, one could make a histogram of the apparent magnitudes of the Pal 5 turnoff stars, shift it by the distance modulus to Pal 5, and obtain the distribution of absolute magnitudes of color-selected Pal 5 turnoff stars.

\citet{2011ApJ...743..187N} showed that the distribution of absolute magnitudes of color-selected turnoff stars were similar for all halo globular clusters observed in the SDSS.  This surprising observation is apparently a result of the age-metallicity relation for the Milky Way galaxy.  Older stellar populations should have fainter, redder turnoffs because a larger number of the more massive stars have had time to evolve away from the main sequence.  On the other hand, the older stellar populations are typically more metal-poor, and metal-poor main sequence stars are brighter and bluer than metal-rich stars of the same mass.  It is surprising, but apparently true, that in the halo of the Milky Way these two effects nearly cancel so that the distribution of absolute magnitudes of blue turnoff stars is essentially the same for all halo stellar populations.  Although the color of the turnoff will be slightly different for different ages and metallicities, the absolute magnitude distributions are surprisingly similar.

This age-metallicity conspiracy provides a great opportunity to apply statistical photometric parallax to measure the spatial density of halo stars, including the substructure from tidal streams.  The observed halo substructure, for example as depicted in Figure~\ref{wedge}, represents the actual density structure, convolved with the absolute magnitude distribution of halo turnoff stars, and then modified by the observational constraints.

To determine the spatial density, we first construct a parameterized spatial density model.  Given a particular set of parameters, one constructs the expected density distribution over a particular volume, derives from this the expected apparent magnitude distribution as a function of angular position in the sky (taking into account the known absolute magnitude distribution of turnoff stars), and then applies observational biases such as completeness.  This expected distribution of stars can then be compared with the observed distribution of stars.  The parameters can be optimized to most closely match the observed data.  For example, \citet{2008ApJ...683..750C} used a maximum likelihood technique that was optimized with conjugate gradient descent to measure the density of tidal streams, and in particular the Sagittarius dwarf tidal stream, in the halo.

As with matched filtering, proper application of this technique includes many subtleties.  For example, if turnoff stars are selected in a narrow color range, then the distribution of stars selected will depend on the color errors.  Typically color errors increase considerably near the observation limits, so that stars with a wider range of colors are sampled at the faint end of the survey.  Because there are few stars bluer than the turnoff, and many intrinsically fainter main sequence stars redder than the turnoff, the distribution of absolute magnitudes of apparently faint stars in a magnitude-limited survey includes a larger number of intrinsically faint stars.

A benefit of the statistical photometric parallax technique is that once the spatial density of a halo substructure is determined, it is possible to {\it statistically} remove stars with the spatial distribution of one halo structure or substructure from the sample, so that the remainder of the sample can be studied.  Note, however, that it is not possible to determine which stars are actually a part of that structure or substructure.

\section{Co-Moving Groups of Stars}

Several halo substructures have been initially identified as group of stars of similar type (RR Lyrae, BHB, etc.), in a small volume, that have similar velocities.  In some cases, these co-moving stars have been associated with a stellar overdensity that is clearly the result of tidal stripping.  In other cases the spatial extent of the related stars is unknown.  In this case the association may be referred to as an Element of Cold Halo Substructure \citep[ECHOS,][]{2009ApJ...703.2177S}.  The disadvantage to this technique is that it requires that the stars have known radial velocities as measured from stellar spectra, or measured proper motions, or both.  The advantage is that as an increasing amount of astrometric and spectroscopic data become available, this technique will allow us to detect and characterize halo substructure with much lower density contrast that result from small mergers, old mergers, mergers where the satellite has been widely spatially dispersed (for example by passing through the center of the Milky Way), or tidal streams in high density regions of the Milky Way.  Kinematically identified halo moving groups are covered in more detail in Chapter 5.

Halo tidal streams have not been identified inside the Solar Circle (within 8 kpc of the center of the Milky Way).  This could be because these more central halo stars formed in situ or because they are due to a large number of well-mixed mergers.  If the central regions formed from mergers, it might be possible to separate the stars into groups with similar angular momentum.  The angular momentum coherence will persist for much longer than spatial coherence, and all of the stars that fall in on a particular satellite have similar angular momentum (see Chapter 6).

\section{Chemical Tagging}

A star's chemical composition is determined by the chemical composition of the gas cloud from which it formed.  Stars in the same open cluster or globular cluster are thought to be coeval and therefore have similar chemical composition (though at least some globular clusters contain more than one stellar population).  If one assumes that most stars are born in star clusters, and that stars maintain a memory of the gas cloud in which they were formed because their outer layers contain the same chemical ratios as those gas clouds, then one could attempt to group stars by the cluster in which they were born \citep{2002ARA&A..40..487F}.  This requires that one make very accurate measurements of a few tens of elemental abundances with high resolution spectroscopy.  The hope is that the chemical abundances of primordial clusters are different enough that they are distinguishable in the observable abundance space.  Chemical tagging, particularly of the Galactic disk, is the primary science driver of the GALAH survey \citep{2015MNRAS.449.2604D}.

In contrast to star clusters, dwarf galaxies generally contain more than one episode of star formation.  While studies of the angular momentum of halo stars will elucidate the history of satellite mergers that built up the halo, studies of stellar chemistry can tell us about the individual star clusters from which the stellar populations are composed.  In particular, the lowest metallicity stars tell us about the conditions in the early Universe, when these stars formed, and about the earliest chemical enrichment processes.  Intermediate metallicity stars require more than one episode of star formation (which is not guaranteed if gas is ejected from low mass dwarf galaxies in the first epoch of star formation), and teach us about the environments in which the stars were formed throughout the evolution of the Universe.  It would be incredible if, at some time in the future, we could build up a record of both the accretion event in which each halo star became a part of the Milky Way, and the star formation event in which that star was born, within the accreted satellite.

Currently, only the simplest chemical tagging techniques have been applied to the detection of tidal streams \citep[for example, the detection of debris from the globular cluster Omega Cen via its unique abundance signatures by][]{2012ApJ...747L..37M}, though streams are often characterized by their chemical composition.

Of particular interest is the variation of chemical composition as a function of the distance from the progenitor.  Because the outer (higher velocity dispersion) layers of a satellite are tidally stripped first, the tidal streams can tell us about the density structure of the progenitor dwarf galaxy.  We expect that higher metallicity stars are are more centrally located in the dwarf galaxy, and therefore will be stripped at later times and be preferentially located nearer to the progenitor dwarf galaxy in the tidal stream. In fact, metallicity gradients have been reported in the Sagittarius dwarf tidal stream (see Chapter 2).

\section{Constraining Dark Matter, the Formation of the Milky Way, and Cosmology with Tidal Streams}
\label{sec:3}

\subsection{Rapid collapse vs. Hierarchical Mergers}
In the late 20$^{\rm th}$ century, there was rapid evolution in our understanding of galaxy formation.  Eggen, Lynden-bell \& Sandage are credited with putting forward the monolithic ``rapid collapse" picture of the formation of the Milky Way, in which halo stars were formed during the collapse of a proto-Galactic gas cloud.  This picture was well summarized in a review by \citet{1973RPPh...36..625E}:
{\it ``The oldest Population II objects (subdwarfs, globular clusters, metal-poor RR Lyrae stars) formed when the galaxy was extended and far from equilibrium, before much metal enrichment occurred. The remainder of the galaxy continued contraction to a disc-like configuration, approximately in centrifugal equilibrium, and dissipated the necessary energy before forming stars; rapid metal enrichment took place. This is the disc population. The residual gas settled to a very flat disc in which the star formation is still going on; this is the Population I."}

\citet{1978ApJ...225..357S} painted a competing picture of galaxy formation, in which they proposed {\it ``that the gas from which the clusters and stars of the outer halo formed continued to fall into the Galaxy for some time after the collapse of its central regions had been completed; that the interactions of the infalling gas dissipated much of its kinetic energy and gave rise to transient high-density regions in which the halo stars and clusters formed; that these regions dispersed even while they underwent chemical evolution; that the stars and clusters that had formed within them eventually fell into dynamical equilibrium with the Galaxy and constitute its present outer halo, while the gas lost from these protogalactic star-forming regions was eventually swept into the galactic disk."}  In other words, the halo stars are stars formed in other, smaller galaxies and clusters, which later merged with a galaxy that had already formed.  This grew into the ``hierarchical merger" scenario, in which the Milky Way grew to its current size through mergers of smaller galaxies and stellar associations.

The hierarchical merger picture has been borne out both by simulations \citep[e.g.][]{1993MNRAS.264..201K,2005Natur.435..629S} and by the fact that we are observing recent and present-day accretion of small galaxies into the Milky Way halo.  Note, however, that the majority of the merging may have happened at early times, and thus the rapid collapse model is not completely incorrect.  Also, hydrodynamical simulations suggest that galactic disks may have formed through steady flow of cold gas into massive dark matter halos \citep{2009Natur.457..451D}.

Because many stars currently in the Milky Way could have been formed in smaller galaxies and later accreted, the star formation history in our galaxy may be different from the history of structure formation.  For example, the composition of stars that were formed outside the Milky Way does not tell us anything about the composition or density of the gas in the Milky Way at the time that star was formed.  There could be an epoch of star formation that is followed by an epoch of merging.

\subsection{Constraining Dark Matter with Tidal Streams}

In the standard $\Lambda$CMD model \citep{spergelLCDM} of the Universe, most of the mass in galaxies like the Milky Way is made of of dark matter, that interacts with itself and the rest of the mass in the Universe only through gravity.  Around Milky Way-sized dark matter halos, there could be thousands of dark matter subhalos that are the size of typical dwarf galaxies.  Because this matter does not interact with light, we cannot see it directly.  Instead we infer its existence from the motions of objects (such as stars and galaxies) that we can see, and from gravitational lensing of objects such as galaxies and QSOs.

Although the stars in the Milky Way halo (often called spheroid stars) represent only 1\% of the stars in the halo, they provide an archaeological record of our galaxy's merger history.  A significant fraction (or maybe all) spheroid stars were formed in other, smaller, galaxies that fell into and were ripped apart by the Milky Way. Any dark matter associated with those galaxies is incorporated into the dark matter halo of the Milky Way.  At least some of the Milky Way globular clusters are believed to have come into the Milky Way with dwarf galaxies that are now disrupted.

The disruption of dwarf galaxies into particular Milky Way tidal streams is typically modeled using $N$-body simulations.  In these simulations, the Milky Way is often assumed to be a fixed potential, and the progenitor satellite is modeled as a symmetric distribution of $N$ (say 10,000 or 100,000) particles of equal mass, whose positions and velocities then evolve through time due to gravitational forces from each other and from the fixed Milky Way potential. Since satellites at 20 to 50 kpc from the Galactic center take of order a billion years to orbit once, these simulations are typically run for billions of years.  However, they cannot be trusted further back in time than about 3 billion years, since the gravitational field of the Milky Way galaxy should evolve with time - particularly at early times in the Universe.  These $N$-body simulations can be used to constrain the properties of the progenitor, the orbit of the satellite from which it was derived, and the gravitational potential of the Milky Way itself.

Alternatively, tidal streams and halo substructure can be modeled in the cosmological context using $N$-body or hydrodynamic simulations of galaxy formation in general.  In this case, the galaxy(ies) and tidal streams will be produced, but will not match the observed galaxies and tidal streams we observe in the Universe today, but they can be used to ask whether the types of structures we see are consistent with the models.  Examples of large simulations of galaxy formation include: the Aquarius Project \citep{2008MNRAS.391.1685S}, Via Lactea \citep{2008Natur.454..735D} and Illustris \citep{2014MNRAS.444.1518V}.

Because the stars in the tidal streams were ripped from their progenitor satellites by the Milky Way's gravity, and their subsequent orbits are determined by the Milky Way's gravity, tidal streams hold great promise for teaching us about the gravitational potential of the Milky Way.  Because the majority of the mass in the Milky Way is believed to be make of dark matter particles, tidal streams could tell us the density distribution of dark matter.

Although in principle one could determine the three dimensional position and velocity of any star in the Milky Way, stars in tidal streams are the only ones that we know where they were in the past.  We know that tidal stream stars were all once co-located on the same progenitor satellite.  This is the property of tidal stream stars that makes them potentially very valuable tracers of the Galactic potential.

Dark matter in the Milky Way galaxy can be described by its overall shape.  In addition to this overall (smoothly varying) density distribution, the radial velocities of the stars in dwarf galaxies suggest that these satellites contain a large fraction of dark matter \citep{2007ApJ...670..313S}. Moreover, cosmological models of structure formation produce galaxies with many more ``subhalos" than there are observed dwarf galaxy satellites; there could be many dwarf galaxy sized lumps of dark matter, that do not contain stars, in orbit around the Milky Way.

Tidal streams have been used to constrain dark matter in all of these contexts (see Chapter 7).  The Sagittarius tidal stream in particular (see Chapter 2) has been used to infer a triaxial shape for the Milky Way's dark matter halo.  Dark matter subhalos can weakly interact with stream stars creating a wider stellar stream, or strongly interact with stream stars, throwing them out of the stream and creating gaps.  The width (or alternatively velocity dispersion) of tidal streams gives some indication of the mass of the progenitor, which can in principle be compared with the number of stars observed to determine the amount of dark matter in a progenitor dwarf galaxy.

\section{Disk Response to Tidal Interactions}

When large dwarf galaxies (or dark matter subhalos) pass close to the Milky Way, the disk can also respond to the gravity of the dwarf galaxy \citep{1969ApJ...155..747H,2006ApJ...641L..33W}.  Historically, this mechanism was studied as the origin of the observed warp in the Galacitic disk, and in particular the warp was thought to be the result of an interaction with the Large Magellanic Cloud (LMC).  After the discovery of the Monoceros Ring (See Chapter 3), disk response to passing satellites was proposed as a mechanism to explain the presence of this highly controversial structure in the plane of the Milky Way \citep{2008ApJ...688..254K,2008ApJ...676L..21Y}.  Later work focussed on the Sagittarius dwarf galaxy, which is known to be close to passing through the disk, as the most likely progenitor \citep{2011Natur.477..301P}.

Surveys of stars have corroborated this picture of disk oscillations.  Coherent disk velocity substructure was observed in the SDSS \citep{2012ApJ...750L..41W}, the RAdial Velocity Experiment \citep[RAVE;][]{2013MNRAS.436..101W}, and the Large Area Multi-Object fiber Spectroscopic Telescope \citep[LAMOST;][]{2013ApJ...777L...5C}.  Oscillations in the disk density have also been observed as a function of height above the plane \citep{2013ApJ...777...91Y}, and distance from the Galactic center \citep{2015ApJ...801..105X}.  The observed density waves are in reasonable agreement with the simulations of \citet{2013MNRAS.429..159G}, who calculated the affects of the Sagittarius dwarf galaxy on the disk.

Evidence for wavelike disk oscillations has also been found in the Andromeda galaxy.  Andromeda's NE clump (see Chapter 8), 40 kpc from the galaxy center, is thought to be composed of disk stars, possibly tossed away from the disk plane by an encounter with the dwarf galaxy progenitor of Andromeda's Giant Stellar Stream \citep{2008AJ....135.1998R,2015MNRAS.446.2789B}.  It is unclear how the corrugations in Galactic gas and star forming regions observed in our galaxy \citep{1974Ap&SS..27..323Q}, and in others \citep{1991MNRAS.251..193F,2001ApJ...550..253A,2008AJ....135..291M}, are related to the disk oscillations that are being observed in Milky Way disk stars, but the subhalos have also been proposed to explain the ripples in the gas \citep{2009MNRAS.399L.118C}.  One wonders whether, if disk oscillations could be caused by galactic subhalos, if they also induce spiral structure.

\section{Future Prospects}

A very dynamic picture of present-day galaxies is emerging.  The stellar halos of large galaxies appear to be composed (in part or in total) of stars that have been tidally stripped from smaller galaxies that have fallen into the larger galaxy's gravitational field.  Massive subhalos (including dark matter-only subhalos, if they exist) could be exciting waves in galactic disks.  We expect that subhalos could produce a time-varying gravitational field in the Milky Way, that could among other things produce gaps in tidal streams.  There is even a possibility that very massive infalling dwarf galaxies could be significantly moving the center of mass of the larger galaxy, affecting not only the disk but also the paths of the dwarf galaxies and associated tidal streams \citep{2015ApJ...802..128G}.

As ongoing and future surveys provide us with large numbers of radial velocities and proper motions of Galactic stars (the largest of these surveys will come from the ESO Gaia satellite), it will be possible to map a larger fraction of halo substructure kinematically (see Chapter 5), and to better constrain the properties of the streams we have already identified.  This better data will allow us to trace tidal stream substructure with higher fidelity, and thus help clarify which stars in tidal streams originated on the same Milky Way satellite.  We will be able to map the spatial extent of kinematically identified streams (which have density contrast too low to follow by density alone), and the kinematic substructure of tidal streams that have been identified by density contrast.

The Sagittarius dwarf tidal stream (see Chapter 2) is presents illuminating example of how far we have come in understanding tidal streams, and how far we still have to go.  This stream is the largest and best studied of the Milky Way tidal streams, but continues to defy a clear explanation of how its structure evolved to match these observations.  We have traced the tidal stream more than $360^\circ$ around the Milky Way in K/M giant, BHB, and turnoff stars.  We have measured the velocities of stream stars and their velocity dispersion, mapped the stream width and the density profile along the stream, measured the chemical compositions and chemical gradients along the stream, compared the properties of the stars in the stream with those in the progenitor galaxy, and identified apparent bifurcations of the stream into two pieces in both the north and south Galactic caps.  However, we are still wondering what combination of Galactic potential and dwarf galaxy progenitor would give rise to these observations.  On the observational side, it is not a certainty that all of the identified parts of the stream are associated with a single dwarf galaxy progenitor.  On the theoretical side, we struggle with how spatially and temporally complex the model for the Galactic potential should be.  We also note that the properties of the progenitor dwarf, including rotation, substructure, satellites, and the radial profiles of the stellar and dwark matter constituents, can give rise to observable complexity in the resulting tidal stream.  On the one hand, tidal streams have the potential to provide us with a wealth of information about our galaxy and the satellites that have fallen into it.  On the other hand, they require us to understand a large number of complexities to reach definite conclusions.

Better understanding of tidal streams will lead to better understanding of galaxy evolution, and to better constraints on the distribution of dark matter, both in the progenitor dwarf galaxies that are thought to have very high mass-to-light ratios, and in the Milky Way galaxy itself.  Techniques for constraining the spatial structure of dark matter are currently in development (See Chapters 6 \& 7).  We still face challenges in understanding the dynamical systems that we are modeling, and in parameterizing the dark matter distribution, that could be quite complex both spatially and in time.  However, the wealth of streams also provide a large number of constraints that one day might unravel the dark matter puzzle.

%
%
%


\begin{thebibliography}{99.}%
\bibitem[Alfaro et al.(2001)]{2001ApJ...550..253A} Alfaro, E.~J., 
P{\'e}rez, E., Gonz{\'a}lez Delgado, R.~M., Martos, M.~A., 
\& Franco, J.\ 2001, ApJ, 550, 253
\bibitem[Belokurov et al.(2006)]{fieldofstreams} V. Belokurov, D. B. Zucker, N. W. Evans, et al., Ap.J. Lett. \textbf{642}, L137 (2006)
\bibitem[Belokurov et al.(2007)]{2007ApJ...657L..89B} Belokurov, V., Evans, N.~W., Bell, E.~F., et al.\ 2007, ApJL, 657, L89 
\bibitem[Bernard et al.(2015)]{2015MNRAS.446.2789B} Bernard, E.~J., 
Ferguson, A.~M.~N., Richardson, J.~C., et al.\ 2015, MNRAS, 446, 2789 
\bibitem[Carlin et al.(2012)]{2012ApJ...753..145C} Carlin, J.~L., Yam, W., Casetti-Dinescu, D.~I., et al.\ 2012, ApJ, 753, 145 
\bibitem[Carlin et al.(2013)]{2013ApJ...777L...5C} Carlin, J.~L., DeLaunay, J., Newberg, H.~J., et al.\ 2013, ApJ Lett, 777, L5 
\bibitem[Chakrabarti \& Blitz(2009)]{2009MNRAS.399L.118C} Chakrabarti, S., \& Blitz, L.\ 2009, MNRAS, 399, L118 
\bibitem[Chiba \& Beers(2000)]{2000AJ....119.2843C} Chiba, M., \& Beers, T.~C., AJ, 119, 2843 (2000)
\bibitem[Cole et al.(2008)]{2008ApJ...683..750C} Cole, N., Newberg, H.~J., Magdon-Ismail, M., et al., ApJ, 683, 750 (2008)
\bibitem[Da Costa \& Armandroff(1995)]{1995AJ....109.2533D} Da Costa, G.~S., \& Armandroff, T.~E., AJ, 109, 2533 (1995)
\bibitem[De Silva et al.(2015)]{2015MNRAS.449.2604D} De Silva, G.~M., 
Freeman, K.~C., Bland-Hawthorn, J., et al.\ 2015, MNRAS, 449, 2604 
\bibitem[Dekel et al.(2009)]{2009Natur.457..451D} Dekel, A., Birnboim, Y., Engel, G., et al.\ 2009, Nature, 457, 451 
\bibitem[Diemand et al.(2008)]{2008Natur.454..735D} Diemand, J., Kuhlen, M., Madau, P., et al.\ 2008, Nature, 454, 735 
\bibitem[Eggen et al.(1973)]{1973RPPh...36..625E} Eggen, O.~J., Freeman, K.~C., \& Rodgers, A.~W., Reports on Progress in Physics, 36, 625 (1973)
\bibitem[Eggen et al.(1962)]{1962ApJ...136..748E} Eggen, O.~J., 
Lynden-Bell, D., \& Sandage, A.~R., Ap. J., 136, 748 (1962)
\bibitem[Eggen \& Sandage(1959)]{1959MNRAS.119..255E} Eggen, O.~J., \& Sandage, A.~R.\ 1959, MNRAS, 119, 255 
\bibitem[Eggen(1958)]{1958MNRAS.118..560E} Eggen, O.~J.\ 1958, MNRAS, 118, 560 
\bibitem[Eggen(1958)]{1958MNRAS.118..154E} Eggen, O.~J.\ 1958, MNRAS, 118, 154 
\bibitem[Eggen(1958)]{1958MNRAS.118...65E} Eggen, O.~J.\ 1958, MNRAS, 118, 65 
\bibitem[Ferguson et al.(2002)]{2002AJ....124.1452F} Ferguson, A.~M.~N., Irwin, M.~J., Ibata, R.~A., Lewis, G.~F., \& Tanvir, N.~R., AJ, 124, 1452 (2002)
\bibitem[Florido et al.(1991)]{1991MNRAS.251..193F} Florido, E., Battaner, E., Sanchez-Saavedra, M.~L., Prieto, M., \& Mediavilla, E.\ 1991, MNRAS, 251, 193 
\bibitem[Freeman \& Bland-Hawthorn(2002)]{2002ARA&A..40..487F} Freeman, K., \& Bland-Hawthorn, J.\ 2002, ARA\&A, 40, 487 
\bibitem[Frenk et al.(1988)]{1988ApJ...327..507F} Frenk, C.~S., White, S.~D.~M., Davis, M., \& Efstathiou, G., ApJ, 327, 507 (1988)
\bibitem[Girard et al.(2004)]{2004AJ....127.3060G} Girard, T.~M., Dinescu, D.~I., van Altena, W.~F., et al., AJ, 127, 3060 (2004)
\bibitem[Helmi et al.(1999)]{Helmi1999} A. Helmi, S. D. M. White, P. T. de Zeeuw, H. Zhao, Nature \textbf{402}, 53 (1999)
\bibitem[Hunter \& Toomre(1969)]{1969ApJ...155..747H} Hunter, C., \& Toomre, A.\ 1969, ApJ, 155, 747 
\bibitem[G{\'o}mez et al.(2013)]{2013MNRAS.429..159G} G{\'o}mez, F.~A., Minchev, I., O'Shea, B.~W., et al.\ 2013, MNRAS, 429, 159 
\bibitem[G{\'o}mez et al.(2015)]{2015ApJ...802..128G} G{\'o}mez, F.~A., Besla, G., Carpintero, D.~D., et al.\ 2015, ApJ, 802, 128 
\bibitem[Grillmair et al.(1995)]{1995AJ....109.2553G} Grillmair, C.~J., Freeman, K.~C., Irwin, M., \& Quinn, P.~J., AJ, 109, 2553 (1995)
\bibitem[Harris(2010)]{2010arXiv1012.3224H} Harris, W.~E.\ 2010, 
arXiv:1012.3224 
\bibitem[Hernquist \& Quinn(1988)]{1988ApJ...331..682H} Hernquist, L., \& Quinn, P.~J.\ 1988, ApJ, 331, 682
\bibitem[Ibata et al.(2003)]{2003MNRAS.340L..21I} Ibata, R.~A., Irwin, 
M.~J., Lewis, G.~F., Ferguson, A.~M.~N., \& Tanvir, N.\ 2003, MNRAS, 340, L21 
\bibitem[Ibata et al.(2001a)]{2001Natur.412...49I} Ibata, R., Irwin, M., Lewis, G., Ferguson, A.~M.~N., \& Tanvir, N., Nature, 412, 49 (2001a)
\bibitem[Ibata et al.(2001b)]{2001ApJ...547L.133I} Ibata, R., Irwin, M., Lewis, G.~F., \& Stolte, A., ApJL, 547, L133 (2001b)
\bibitem[Ibata et al.(1994)]{1994Natur.370..194I} Ibata, R.~A., Gilmore, G., \& Irwin, M.~J., Nature, 370, 194 (1994)
\bibitem[Ibata et al.(1997)]{1997AJ....113..634I} Ibata, R.~A., Wyse, 
R.~F.~G., Gilmore, G., Irwin, M.~J., \& Suntzeff, N.~B., AJ, 113, 634 (1997)
\bibitem[Ivezi{\'c} et al.(2000)]{2000AJ....120..963I} Ivezi{\'c}, {\v Z}., Goldston, J., Finlator, K., et al., AJ, 120, 963 (2000)
\bibitem[Ivezi{\'c} et al.(2008)]{2008ApJ...684..287I} Ivezi{\'c}, {\v Z}., Sesar, B., Juri{\'c}, M., et al., ApJ, 684, 287 (2008)
\bibitem[Ivezi{\'c} et al.(2012)]{2012ARAA..50..251I} Ivezi{\'c}, {\v Z}., Beers, T.~C., \& Juri{\'c}, M.\ 2012, ARA\&A, 50, 251 
\bibitem[Johnston et al.(1995)]{1995ApJ...451..598J} Johnston, K.~V., 
Spergel, D.~N., \& Hernquist, L., ApJ, 451, 598 (1995)
\bibitem[Johnston et al.(1996)]{1996ApJ...465..278J} Johnston, K.~V., 
Hernquist, L., \& Bolte, M., ApJ, 465, 278 (1996)
\bibitem[Juri{\'c} et al.(2008)]{2008ApJ...673..864J} Juri{\'c}, M., 
Ivezi{\'c}, {\v Z}., Brooks, A., et al., ApJ, 673, 864 (2008)
\bibitem[Kauffmann et al.(1993)]{1993MNRAS.264..201K} Kauffmann, G., White, S.~D.~M., \& Guiderdoni, B., MNRAS, 264, 201 (1993)
\bibitem[Kazantzidis et al.(2008)]{2008ApJ...688..254K} Kazantzidis, S., Bullock, J.~S., Zentner, A.~R., Kravtsov, A.~V., \& Moustakas, L.~A.\ 2008, ApJ, 688, 254 \bibitem[Kepley et al.(2007)]{2007AJ....134.1579K} Kepley, A.~A., Morrison, H.~L., Helmi, A., et al.\ 2007, AJ, 134, 1579 
\bibitem[Lenz et al.(1998)]{1998ApJS..119..121L} Lenz, D.~D., Newberg, J., Rosner, R., Richards, G.~T., \& Stoughton, C.\ 1998, ApJ Suppl, 119, 121 
\bibitem[Lynden-Bell \& Lynden-Bell(1995)]{1995MNRAS.275..429L} Lynden-Bell, D., \& Lynden-Bell, R.~M.\ 1995, MNRAS, 275, 429 
\bibitem[Majewski et al.(2012)]{2012ApJ...747L..37M} Majewski, S.~R., 
Nidever, D.~L., Smith, V.~V., et al.\ 2012, ApJ Lett, 747, L37 
\bibitem[Majewski et al.(2003)]{2003ApJ...599.1082M} Majewski, S.~R., 
Skrutskie, M.~F., Weinberg, M.~D., \& Ostheimer, J.~C., ApJ, 599, 1082 (2003)
\bibitem[Majewski et al.(2000)]{2000AJ....120.2550M} Majewski, S.~R., 
Ostheimer, J.~C., Kunkel, W.~E., \& Patterson, R.~J.\ 2000, AJ, 120, 2550 
\bibitem[Majewski et al.(1996)]{mmh1996} S. R. Majewski, J. A. Munn, S. L. Hawley, Ap.J. Lett. \textbf{459}, L73 (1996)
\bibitem[Matthews \& Uson(2008)]{2008AJ....135..291M} Matthews, L.~D., \& Uson, J.~M.\ 2008, AJ, 135, 291 
\bibitem[McConnachie(2012)]{2012AJ....144....4M} McConnachie, A.~W.\ 2012, AJ, 144, 4 
\bibitem[Morrison et al.(2000)]{2000AJ....119.2254M} Morrison, H.~L., 
Mateo, M., Olszewski, E.~W., et al., AJ, 119, 2254 (2000)
\bibitem[Newberg et al.(2002)]{NYetal2002} H. J. Newberg, B. Yanny, C. Rockosi, et al., Ap.J. \textbf{569}, 245 (2002)
\bibitem[Newberg et al.(2009)]{2009ApJ...700L..61N} Newberg, H.~J., Yanny, B., \& Willett, B.~A.\ 2009, ApJL, 700, L61 
\bibitem[Newberg(2013)]{2013IAUS..289...74N} Newberg, H.~J., IAU Symposium, 289, 74 (2013)
\bibitem[Newby et al.(2011)]{2011ApJ...743..187N} Newby, M., Newberg, 
H.~J., Simones, J., Cole, N., \& Monaco, M., ApJ, 743, 187 (2011)
\bibitem[Newby et al.(2013)]{2013AJ....145..163N} Newby, M., Cole, N., Newberg, H.~J., et al., AJ, 145, 163 (2013)
\bibitem[Odenkirchen et al.(2001)]{2001ApJ...548L.165O} Odenkirchen, M., Grebel, E.~K., Rockosi, C.~M., et al., ApJL, 548, L165 (2001)
\bibitem[Purcell et al.(2011)]{2011Natur.477..301P} Purcell, C.~W., 
Bullock, J.~S., Tollerud, E.~J., Rocha, M., \& Chakrabarti, S.\ 2011, Nature, 477, 301 
\bibitem[Quiroga(1974)]{1974Ap&SS..27..323Q} Quiroga, R.~J.\ 1974, Ap\&SS, 27, 323
\bibitem[Richardson et al.(2008)]{2008AJ....135.1998R} Richardson, J.~C., Ferguson, A.~M.~N., Johnson, R.~A., et al.\ 2008, AJ, 135, 1998 
\bibitem[Rockosi et al.(2002)]{2002AJ....124..349R} Rockosi, C.~M., 
Odenkirchen, M., Grebel, E.~K., et al.\ 2002, AJ, 124, 349 
\bibitem[Rodgers et al.(1981)]{1981ApJ...244..912R} Rodgers, A.~W., 
Harding, P., \& Sadler, E., ApJ, 244, 912 (1981)
\bibitem[Searle \& Zinn(1978)]{1978ApJ...225..357S} Searle, L., \& Zinn, R., Ap. J., 225, 357 (1978)
\bibitem[Schlaufman et al.(2009)]{2009ApJ...703.2177S} Schlaufman, K.~C., Rockosi, C.~M., Allende Prieto, C., et al.\ 2009, ApJ, 703, 2177 
\bibitem[Shang et al.(1998)]{1998ApJ...504L..23S} Shang, Z., Zheng, Z., Brinks, E., et al.\ 1998, ApJ Lett, 504, L23 
\bibitem[Sharma et al.(2010)]{2010ApJ...722..750S} Sharma, S., Johnston, K.~V., Majewski, S.~R., et al.\ 2010, ApJ, 722, 750 
\bibitem[Simon \& Geha(2007)]{2007ApJ...670..313S} Simon, J.~D., \& Geha, M.\ 2007, ApJ, 670, 313 
\bibitem[Spergel et al.(2015)]{spergelLCDM} Spergel, D. N., Science, 347, 1100 (2015)
\bibitem[Springel et al.(2008)]{2008MNRAS.391.1685S} Springel, V., Wang, J., Vogelsberger, M., et al.\ 2008, MNRAS, 391, 1685 
\bibitem[Springel et al.(2005)]{2005Natur.435..629S} Springel, V., White, S.~D.~M., Jenkins, A., et al., Nature, 435, 629 (2005)
\bibitem[Starkenburg et al.(2009)]{2009ApJ...698..567S} Starkenburg, E., Helmi, A., Morrison, H.~L., et al., ApJ, 698, 567 (2009)
\bibitem[Vivas et al.(2001)]{2001ApJ...554L..33V} Vivas, A.~K., Zinn, R., Andrews, P., et al., ApJL, 554, L33 (2001)
\bibitem[Vogelsberger et al.(2014)]{2014MNRAS.444.1518V} Vogelsberger, M., Genel, S., Springel, V., et al.\ 2014, MNRAS, 444, 1518 
\bibitem[Weinberg \& Blitz(2006)]{2006ApJ...641L..33W} Weinberg, M.~D., \& Blitz, L.\ 2006, ApJ Lett, 641, L33 
\bibitem[Widrow et al.(2012)]{2012ApJ...750L..41W} Widrow, L.~M., Gardner, S., Yanny, B., Dodelson, S., \& Chen, H.-Y.\ 2012, ApJ Lett, 750, L41 
\bibitem[Williams et al.(2013)]{2013MNRAS.436..101W} Williams, M.~E.~K., Steinmetz, M., Binney, J., et al.\ 2013, MNRAS, 436, 101 
\bibitem[Xu et al.(2015)]{2015ApJ...801..105X} Xu, Y., Newberg, H.~J., 
Carlin, J.~L., et al.\ 2015, ApJ, 801, 105 
\bibitem[Yanny et al.(2000)]{2000ApJ...540..825Y} Yanny, B., Newberg, 
H.~J., Kent, S., et al., ApJ, 540, 825 (2000)
\bibitem[Yanny \& Gardner(2013)]{2013ApJ...777...91Y} Yanny, B., \& Gardner, S.\ 2013, ApJ, 777, 91 
\bibitem[York et al.(2000)]{2000AJ....120.1579Y} York, D.~G., Adelman, J., Anderson, J.~E., Jr., et al., AJ, 120, 1579 (2000)
\bibitem[Younger et al.(2008)]{2008ApJ...676L..21Y} Younger, J.~D., Besla, G., Cox, T.~J., et al.\ 2008, ApJ Lett, 676, L21 

\end{thebibliography}
\end{document}